# DYNAMICS OF THE GALACTIC GLOBULAR CLUSTER NGC 3201

**Patrick Côté**[1,2]

Department of Physics and Astronomy, McMaster University

Hamilton, ON L8S 4M1, Canada

and

Dominion Astrophysical Observatory, Herzberg Institute of Astrophysics

National Research Council, 5071 West Saanich Road, Victoria, BC, V8X 4M6, Canada

cote@dao.nrc.ca

**Douglas L. Welch**[1,2]

Department of Physics and Astronomy, McMaster University

Hamilton, ON L8S 4M1, Canada

welch@physun.physics.mcmaster.ca

**Philippe Fischer**[1,2]

Department of Physics and Astronomy, McMaster University

Hamilton, ON L8S 4M1, Canada

and

AT&T Bell Laboratories, 600 Mountain Avenue, 1D-316, Murray Hill, NJ 07974

philf@physics.att.com

**K. Gebhardt**[3]

Department of Physics and Astronomy, Rutgers University, Piscataway, NJ 08855-0849

gebhardt@astro.lsa.umich.edu

[1] Visiting Astronomer at the Cerro Tololo Inter-American Observatory, National Optical Astronomy Observatories, operated by the Association of Universities for Research in Astronomy, Inc., under contract with the National Science Foundation.

[2] Guest Investigator, Mount Wilson and Las Campanas Observatories, which are operated by the Carnegie Institute of Washington.

[3] Present Address: Department of Astronomy, University of Michigan, Dennison Building, Ann Arbor, MI 48109-1090, USA

## ABSTRACT

*BV* CCD frames have been used to derive surface brightness profiles for NGC 3201 which extend out to approximately 18′. A total of 857 radial velocities with median precision $\simeq$ 1 km s$^{-1}$ for 399 member giants have been used to trace the velocity dispersion profile out to 32.1′ (the approximate tidal radius determined from fits of single-mass, isotropic King-Michie models to the cluster surface brightness profiles). The median difference in radial velocity for stars on either side of an imaginary axis stepped through the cluster in 1° increments shows a statistically significant maximum amplitude of 1.22±0.25 km s$^{-1}$. We discuss several possible explanations of this result, including: (1) cluster rotation; (2) preferential stripping of stars on prograde orbits near the limiting radius; (3) the projection of the cluster space velocity onto the plane of the sky and; (4) a slight drift in the velocity zero point. It is difficult to unambiguously identify the primary cause of the observed structure in the velocity field, however, and we suspect that all of the above processes may play a role. The *BV* surface brightness profiles and radial velocities have been modeled with both single- and multi-mass King-Michie models and nonparametric techniques. The corresponding density- and M/L-profiles show good agreement over the interval 1.5 $\lesssim$ R $\lesssim$ 10 pc, and both approaches suggest a steady rise in M/L with distance from the cluster center. Due to the low cluster luminosity, we are unable to place useful constraints on the anisotropy of the velocity dispersion profile, though the global mass-to-light ratio is well-constrained by the models: $M/L_B \simeq M/L_V \simeq 2.0\pm0.2$ for the multi-mass and nonparametric models, compared to $\simeq$ 1.65±0.15 for models having equal-mass stars. Our best-fit, multi-mass models have mass function slopes of $x \simeq 0.75 \pm 0.25$, consistent with recent findings that the form of the mass function depends on the position relative to the potential of the Galaxy.

## 1. INTRODUCTION

Improved observational constraints on the internal dynamics of globular clusters are demanded by many of the most fundamental questions regarding their formation and evolution. For instance, does the velocity dispersion profile (VDP) fall off with projected distance from the cluster center in the manner predicted by multi-mass, King-Michie models (Da Costa and Freeman 1976; Gunn and Griffin 1979) or does an appreciable amount of dark matter reside in the envelope of some clusters, giving rise to a flat VDP? Do global M/Ls vary from cluster to cluster and how does the M/L change with radius in a given cluster? What is the form of the cluster mass function and how significant are the observed correlations of mass function slope with cluster position in the Galaxy (Capaccioli, Piotto and Stiavelli 1993)? How abundant are primordial binaries in these Population II systems (Hut *et al.* 1992; Côté *et al.* 1994) and what is their radial distribution? How common are central velocity dispersion cusps (Peterson, Seitzer and Cudworth 1989) and do they reflect post core-collapse evolution (Spitzer 1985; Grabhorn *et al.* 1992) or the presence of massive central bodies (Newell, Da Costa and Norris 1976)? Are stellar orbits in the outer regions of the cluster predominantly radial or has the tidal field of the Galaxy induced isotropy near the tidal radius, as suggested by the three-body/Fokker-Planck models of Oh and Lin (1992)? And to what extent do the underlying dynamics affect the mix of stellar populations (see Trimble and Leonard 1994 for a recent review)? Clearly, answers to many of these questions require an understanding of how the velocity dispersion varies from the cluster core to the tidal radius.

Early measurements of globular cluster velocity dispersions were based on the broadening of stellar absorption lines in long-slit spectra of the integrated cluster light (Illingworth 1976). However, since the requisite measurements are possible only in the cluster core (and are complicated by the presence of central binaries which tend to produce overestimates of the dispersion and luminous giants which often dominate the measured spectrum; Zaggia *et al.* 1992), this technique is capable of providing little more than a central M/L for a given cluster. An alternative approach is to use proper motions of individual stars to determine the run of velocity dispersion. Though potentially very powerful (these observations contain the two components of the VDP needed to solve the non-rotating Jeans equation; Leonard *et al.* 1992), proper motions of the requisite precision are exceedingly difficult to measure for stars in such crowded fields (*e.g.* Cudworth and Monet 1979).

Most work on cluster dynamics has therefore made use of individual stellar radial velocities, since a precision of $\simeq 1$ km s$^{-1}$ is often attainable using large telescopes, high-QE detectors and cross-correlation techniques. Nevertheless, progress has been slow, since the measurement of a hundred or more such velocities with a single-channel spectrograph or a radial velocity scanner is tremendously time-consuming. As a consequence, only a handful of dynamical studies based on large radial velocity samples (N $\gtrsim$ 100) have appeared in print (*e.g.*, M3, Gunn and Griffin 1979; M2, Pryor *et al.* 1986; $\omega$ Cen and 47 Tuc, Meylan and Mayor 1986; M13, Lupton, Gunn and Griffin 1987; NGC 6397, Meylan, Mayor and Dubath 1991 and NGC 362, Fischer *et al.* 1993). In addition, the sequential nature of the observations has restricted work (with one notable exception; Seitzer 1983) primarily to the inner cluster regions where the probability of observing member stars is highest. Once radial velocities are in hand, cluster membership is more easily established, though for many clusters, the velocity-space distributions of field and cluster stars show considerable overlap. In these cases, even with kinematic information, assigning cluster membership remains a rather dubious business.

With the introduction of multi-object spectrographs on many 4.0m-class telescopes, surveys to trace VDPs over the full range in cluster radius have become feasible. In this paper, we present a dynamical study of the Galactic globular cluster NGC 3201 based on 857 radial velocities for 399 member stars which have been used to derive a projected VDP which extends from the core to the approximate tidal radius. NGC 3201 is the logical cluster for such an endeavor, since its systemic radial velocity of 494 km s$^{-1}$ ensures no overlap with the field star population (see Table 1 for a summary of general cluster properties). The radial velocities used in this analysis have been presented in a companion paper (Côté *et al.* 1994) and were accumulated primarily with ARGUS, the fiber-fed, bench-mounted, multi-object spectrograph on the CTIO 4.0m telescope. ARGUS is ideally suited for a complete sampling of the VDP since it offers (1) high velocity precision with the echelle grating, (2) the ability to acquire spectra for 24 stars simultaneously, (3) a minimum fiber separation of $10''$ (an important consideration for the crowded cores of most globular clusters) and (4) a $50'$ field of view which allows the simultaneous observation of both core and envelope stars. The resulting VDP has been combined with $BV$ surface brightness profiles (SBPs) based on CCD photometry to investigate the cluster dynamics using both single- and multi-mass King-Michie models (Michie 1963; King 1966a; Da Costa and Freeman 1976; Gunn and Griffin 1979) and nonparametric models (Merritt and Tremblay 1994; Gebhardt and Fischer 1995).

## 2. OBSERVATIONS AND REDUCTIONS

Realistic models of globular clusters require a knowledge of not only the light profile but also the radial variation in velocity dispersion (see Lupton, Gunn and Griffin 1985). In this section we describe the data upon which our SBPs and VDP for NGC 3201 are based.

### 2.1 *Surface Photometry and Star Counts*

SBPs for NGC 3201 were constructed from $BV$ CCD frames collected with the 1.0m Swope telescope at Las Campanas Observatory on 21/22 January and 22/23 February 1991. The detector used was the 1024×1024 Tek2 CCD (readnoise = $7e^-$, gain = $2e^-$/ADU and scale = $0.609''$/pixel) so that each image measures $10.4' \times 10.4'$. Exposure times were 120s for $V$ and 180s for $B$. $BV$ frame pairs were obtained for five separate fields on the night on 22/23 January 1991: one centered on the cluster core and four offset by roughly $7'$ toward the NE, SE, SW, and NW directions (Figure 1 shows the relative positioning of these five fields). Another sequence of $BV$ frames was obtained on the night of 22/23 February 1991, this time for fields offset from the cluster center by $\sim 14'$, $23'$, $32'$, $41'$ and $50'$ in both the N and S directions. Frames were bias-subtracted, overscan-corrected, trimmed and flat-fielded with the usual IRAF[4] tasks. Instrumental magnitudes were determined with DoPHOT (Schechter *et al.* 1993) and calibrated using nine unsaturated, on-frame photoelectric standards chosen from the lists of Alcaino and Liller (1984) and Lee (1977). A comparison of our photometry with that of Brewer *et al.* (1993) showed excellent agreement in $V$ and a slight (but systematic) difference in $(B-V)$ in the sense that our inferred colors are $\simeq$ 0.04 mag redder than those of Brewer *et al.* (1993).

Due to its low Galactic latitude (see Table 1) and the fact that it is a sparse cluster (concentration class = X; Shapley 1930), determining a reliable background level for NGC 3201 is somewhat problematic. Previous work based on visual star counts made on photographic plates (Peterson and King 1975; King *et al.* 1968) placed the cluster tidal radius at $r_t \simeq 36'$. We therefore used our DoPHOT photometry for all fields out to $\simeq 50'$ to perform stars counts in concentric annuli positioned on the cluster center found by Shawl and White (1986). Only main-sequence turnoff stars and evolved giants were used to construct the surface density profiles, ensuring that the measured SBPs correspond to stars of almost identical mass. Of course, crowding in the cluster core reduces the completeness of the star counts — in this region we performed surface photometry in the manner described by Fischer *et al.* (1993). The central CCD images were divided into concentric annuli positioned on the cluster center. These annuli were then divided into eight azimuthal sections; the mean pixel value for each of these sectors was then determined and the median of these eight measurements was adopted as the surface brightness for the annulus (at the area-weighted mean radius). The uncertainty in the surface photometry was taken to be the standard error in the median of the eight sectors; Poisson statistics were used to determine the corresponding uncertainties in the star counts. The surface photometry and star count surface densities were then merged by matching (via least-squares) the two datasets in the range $4 \lesssim \mathrm{R} \lesssim 9$ pc (where incompleteness in the star counts was negligible). The final, background-subtracted $BV$ SBPs for NGC 3201 are given in Table 2 which records the projected radius, the adopted surface brightness and its source. In converting R and $\mu$ to pc and $L_\odot$ pc$^{-2}$, we have adopted $M_{V,\odot} = 4.83$, $M_{B,\odot} = 5.48$ (Binney and Tremaine 1987), an apparent distance modulus of $(m-M)_V = 14.20 \pm 0.15$ and a cluster reddening of $E(B-V) = 0.21 \pm 0.02$ (Brewer *et al.* 1993) so that 1pc = 40.24$''$ at NGC 3201.

[4] IRAF is distributed by the National Optical Astronomy Observatories, which are operated by the Association of Universities for Research in Astronomy, Inc., under contract to the National Science Foundation.

### 2.2 *Radial Velocities*

As previously noted, the number of dynamical studies of globular clusters based on large radial velocity samples is rather small. Moreover, the observed VDPs for these clusters generally extend to only $r/r_t \approx 0.25$. Our reasons for observing cluster members at large projected radii in NGC 3201 were twofold: (1) to trace the cluster VDP out to $\simeq r_t$ and (2) to search for primordial binaries and find their radial distribution. The results of our search for binaries in NGC 3201 have already been published, along with the entire sample of NGC 3201 radial velocities (Côté *et al.* 1994).[5] The reader is referred to the above reference for a more complete discussion of the spectroscopic observations and reductions.

Spectra were accumulated during several observing runs with telescopes at both Las Campanas and CTIO. Photon-counting echelle spectrographs on the Las Campanas 2.5m and CTIO 4.0m telescopes were used to measure 267 radial velocities for 189 stars (chosen from the finder charts of Lee 1977) within $\simeq 5'$ of the cluster center during observing runs in January/February 1991. Object spectra in the range 5120 – 5460 Å were cross-correlated against template spectra for a variety of IAU radial velocity standard stars to give heliocentric radial velocities with precision 1.3 – 1.7 km s$^{-1}$. The bulk of the spectra were obtained during two observing runs (February 15 – 16 and March 15 – 16 1992) with ARGUS: the bench-mounted, fiber-fed, multi-object spectrograph on the CTIO 4.0m telescope. Object spectra in the range 5090 – 5160 Å were cross-correlated against high S/N spectra of the twilight/dawn sky. Repeat observations suggest that the ARGUS velocities have a median accuracy of $\sim 1$ km s$^{-1}$. This sample of 1730 radial velocities for 1316 stars was then combined with the 129 radial velocities (92 member stars) used in the lone previous dynamical study of NGC 3201 (Da Costa *et al.* 1993) and also published in Côté *et al.* (1994).

[5] Table 3 of Côté *et al.* 1994 is also available on the AAS CD-ROM Series, Volume 5, 1995.

The complete survey therefore consists of 1859 radial velocities for 1318 stars within $36'$ of the cluster center. As pointed out earlier, the high systemic radial velocity of 494 km s$^{-1}$ for NGC 3201 (see Figure 1 of Côté *et al.* 1994 and below) ensures the unambiguous identification of all field stars (a total of 889 radial velocities for 879 field stars were accumulated). Tables 2 and 3 of Côté *et al.* (1994) list all 970 radial velocities for 439 cluster members. Any radial velocity variables in the sample such as RR Lyraes or binary stars will lead to overestimates of the velocity dispersion and must be removed from the final sample; the 19 known photometric variables in our survey (Fourcade and Laborde 1966; Sawyer-Hogg 1973) were therefore omitted along with the 21 candidate binaries listed in Côté *et al.* (1994). The final sample therefore consists of (weighted) mean velocities for 399 cluster members in the range $0.08' \leq$ R $\leq 32.1'$ based on 857 radial velocities. Absolute positions with precision $\sim 1''$ for all 399 program stars, derived from our CCD frames and APM scans using the HST Guide Star Catalog, are recorded in Côté *et al.* (1994).

Estimates of the mean velocity $v_0$ and intrinsic velocity dispersion $\sigma_0$ of NGC 3201 can be obtained with the formulae of Armandroff and Da Costa (1986). However, as noted by Suntzeff *et al.* (1993), care must be taken in applying these forumlae since the Armandroff and Da Costa (1986) estimator of the error in the intrinsic variance assumes an *unweighted* variance, not the *weighted* variance given by their formulae. We have therefore followed the prescription of Suntzeff *et al.* (1993) in deriving the mean velocity and intrisic velocity dispersion, although for the present sample the difference amounts to less than a few percent. Based on the above sample of 420 velocities (*i.e.* excluding only the known photometric variables), we find $v_0 = 494.0 \pm 0.2$ km s$^{-1}$ and $\sigma_0 = 3.70 \pm 0.13$ km s$^{-1}$.

Removing the 21 binary candidates listed in Côté *et al.* (1994) changes these numbers only slightly: $v_0 = 494.0 \pm 0.2$ km s$^{-1}$ and $\sigma_0 = 3.66 \pm 0.13$ km s$^{-1}$. Considering only the 93 stars within $1.46'$ (*i.e.* one core radius) of the cluster center gives $v_0 = 493.2 \pm 0.4$ km s$^{-1}$ and $\sigma_0 = 3.88 \pm 0.28$ km s$^{-1}$. For all three samples, these estimates of $v_0$ and $\sigma_0$ are virtually identical to those obtained using the technique of Peterson and Latham (1986). Using the maximum-likelihood approach of Pryor and Meylan (1993) yields $v_0 = 494.4 \pm 0.2$ km s$^{-1}$ and $\sigma_0 = 3.77 \pm 0.16$ km s$^{-1}$ for all 420 stars, $v_0 = 494.4 \pm 0.2$ km s$^{-1}$ and $\sigma_0 = 3.69 \pm 0.13$ km s$^{-1}$ for the restricted sample of 399 stars and, $v_0 = 494.5 \pm 0.4$ km s$^{-1}$ and $\sigma_0 = 3.64 \pm 0.25$ km s$^{-1}$ for the 93 stars within one core radius. In §3.1.2, we review the maximum-likelihood estimators of the systemic velocity and velocity dispersion devised by Gunn and Griffin (1979). Although these estimates are model-dependent, we find $v_0 \sim 494.2$ km s$^{-1}$ and $\sigma_0 \sim 4.3$ km s$^{-1}$ using this approach, in good agreement with the above results (note that both the nonparametric and binned VDPs show central dispersions which are slightly *lower* than that seen at intermediate radii; the maximum-likelihood scaling of the single- and multi-mass model VDPs makes use of *all* of the velocities and therefore leads to slightly larger estimates for the central dispersion). The dispersion profile is discussed in more detail in §3.1.

### 2.3 *Possible Structure in the Velocity Field*

The relationship between heliocentric radial velocity and both radius and position angle is shown in the upper and middle panels of Figure 2. The latter of these plots suggests some dependence of the observed velocity on position angle, a trend which is more apparent in the lower panel of Figure 2, where we have plotted the median radial velocity versus position angle for eight azimuthal bins of equal width. The results are summarized in Table 3 which records the bin number, the number of stars in each sector, the range in position angle, the mean position angle for the sector and the median radial velocity. Another way of identifying an azimuthal dependence of radial velocity, and one commonly used to search for rotation in globular clusters, is to step an imaginary axis through the cluster in small, angular increments and compute the median velocity difference on either side of this line, $\Delta V_{r,\rm med}$, for each angle. The results of such a procedure are shown in the upper panel of Figure 3 which shows the dependence of $\Delta V_{r,\rm med}$ on axis angle $\Phi$. Also shown is the best-fit sine curve which has an amplitude of $1.22 \pm 0.25$ km s$^{-1}$ and a phase shift of $277 \pm 12^\circ$, implying a position angle for the axis of $\Phi = -7 \pm 12^\circ$.

How significant is this detection? To answer this question, we have generated 1000 artificial datasets (*i.e.* 399 radial velocities at the corresponding locations of our program stars). The simulated radial velocity for each star has been chosen by adopting the mean dispersion for a star at the projected radius of the program object (estimated from the best-fit, single-mass, isotropic King-Michie model). For each star, a realistic amount of observational noise (typical velocity uncertainty $\simeq 1$ km s$^{-1}$) has been included. Each simulated dataset was then analyzed in a manner identical to that used for the real data. The histogram of the resulting amplitudes is given in the lower panel of Figure 3. Only eight times in 1000 trials did the best-fit sine wave have an amplitude of 1.22 km s$^{-1}$ or greater; we therefore conclude that the observed signal is significant at the 99.2% level.

Of course, such simulations do not account for possible differences in the velocity zero points from different runs. For example, since the bulk of the radial velocities were accumulated during a pair of two-night observing runs with Argus, it is possible that a drift in the velocity zero point could give rise to the observed trend, *provided the field of view studied on a given night is appreciably smaller than the total field of view.* As discussed in Côté *et al.* (1994), small zero point corrections were applied to the velocities accumulated during different observing runs in order to bring them onto a common system. As a result, the mean velocities for two Argus runs show good agreement: 494.38 km s$^{-1}$ and 494.37 km s$^{-1}$ for the first and second runs, respectively. On the other hand, the mean velocities for the first and second nights of the second Argus run show an offset of 1.4 km s$^{-1}$, suggesting that a shift in the velocity zero point may be to blame. However, it is unlikely that such a shift is *solely* responsible for the observed trend, since the sample of stars observed on March 15/16 1992 and March 16/17 1992 have almost indentical distributions with respect to the total field of view of the survey.

We now discuss a number of other possible origins of the observed dependence of radial velocity on position angle: cluster rotation, the stripping of stars near the tidal radius by encounters with the Galactic disk and the projection of the cluster space velocity onto the plane of the sky.

### 2.3.1 *Rotation*

Since the lone previous dynamical study of NGC 3201 (based on mean radial velocities for 92 stars; Da Costa *et al.* 1993) found appreciable rotation in the range $1.3' \leq r \leq 3.2'$ where 51 stars showed a formally significant rotation amplitude of 0.7±0.2 km $s^{-1}$, it would not be surprising if a small amount of rotation was observed in our sample of velocities. It is therefore natural to ask whether or not the velocity difference of 1.22 km $s^{-1}$ evident in Figure 3 can be due solely to rotation. If we assume that the observed amplitude is, in fact, purely a consequence of cluster rotation, we have $V_{\rm rot}/\sigma \simeq 1.22/3.67 = 0.33 \pm 0.08$ for the ratio of ordered ($V_{\rm rot}$) to random ($\sigma$) motions, consistent with the theoretical ratio for the purely rotationally-flattened case.[6] The location of NGC 3201 in the $\epsilon$-$V_{\rm rot}/\sigma$ plane is given in Figure 4. For comparison, we also show the ($\epsilon$, $V_{\rm rot}/\sigma$)-relation for rotationally-flattened oblate spheroids with isotropic velocity dispersion tensors (Binney 1978; Binney and Tremaine 1987). Further evidence that the observed velocity difference is at least partly caused by rotation is provided by the orientation of the axis which maximizes $\Delta V_{r,\rm med}$ — the kinematically determined rotation axis is located at a (projected) position angle of $-7 \pm 12°$, in excellent agreement with the position angle of the cluster's photometric minor axis ($2 \pm 7°$ according to White and Shawl 1987). That is, both the amplitude and position of the NGC 3201 rotation axis are in good agreement with that expected for a rotationally-flattened oblate spheroid. Nevertheless, we believe that it is unlikely that rotation *alone* is the cause of the observed dependence on position angle. Although the good agreement between (1) the assumed rotation axis and the photometric minor axis and (2) the observed and expected cluster ellipticity lends support to the notion that rotation is partly responsible for observed trend in velocity, the amplitudes of the best-fitting sine-curves fit to increasingly distant samples of radial velocities show an unexpected *increase* with radius (see Table 4), suggesting that some other effect may also be at work.

[6] Strictly speaking, $V_{\rm rot}$ and $\sigma$ are the projected, mass-weighted rotation velocity and the projected, mass-weighted velocity dispersion (the computation of which require the adoption of a rotation model). While the second approximation is generally a good one, this procedure will tend to overestimate $V_{\rm rot}$ since the true rotation curve probably peaks away from the cluster center; it is therefore probably best to view the resulting value of $V_{\rm rot}/\sigma$ as an upper limit.

### 2.3.2 *Tidal Stripping*

An increase in apparent rotation at large radii has been predicted by Oh and Lin (1992), who carried out an investigation of the tidal evolution of globular clusters using a Fokker Planck/three-body integration approach. They confirmed earlier findings (Keenan and Innanen 1975; Jefferys 1976; Keenan 1981) that stars on direct orbits are less stable than their retrograde counterparts. Prolonged interaction with the Galactic tidal field therefore results in preferential stripping of such stars and can lead to an apparent rotation of the cluster. Oh and Lin (1992) also note that such an apparent rotation can be extended into relatively small radii for clusters with appreciable velocity anisotropy — not inconsistent with the results of our dynamical modeling (see §4.4). It is therefore possible that such a process is at work in NGC 3201 and has contributed to the apparent cluster rotation at large radii.

### 2.3.3 *Motion Across the Line of Sight*

Finally, we note that another, albeit more speculative, explanation of the observed trend is possible: if NGC 3201 has a substantial component of its systemic velocity directed across the line of sight, then the observed dependence of radial velocity on position relative to the cluster core may be a result of slightly different projections of the cluster space velocity along the line of sight (since the radial velocities are scattered over an area of nearly one square degree; see Figure 5). If it is assumed that the velocity variations are solely due to systemic motion across the line of sight, *the cluster space velocity can be computed using only the observed radial velocities.*

In a rectangular coordinate system with $X$ toward $\alpha = 0°$, $\delta = 0°$, $Y$ toward $\alpha = 90°$, $\delta = 0°$ and $Z$ toward the north celestial pole, we can neglect the radial velocity dispersion of the cluster and write (Feast *et al.* 1961)

$$v_i = X \cos\alpha_i \cos\delta_i + Y \sin\alpha_i \cos\delta_i + Z \sin\delta_i \tag{1}$$

where $v_i$ is the observed radial velocity of a cluster member, $\alpha_i, \delta_i$ are its coordinates and $XYZ$ are the components of the cluster space velocity. The velocity components which minimize the $\chi^2$ of the above equation are: $X = -409.9 \pm 25.0$ km s$^{-1}$, $Y = -23.0 \pm 23.2$ km s$^{-1}$ and $Z = -340.0 \pm 23.2$ km s$^{-1}$. All 399 cluster members have been used in the fit, with the radial velocities weighted by $\sigma_{i,\mathrm{w}} = ({\sigma_i}^2 + {v_s}^2{\eta_i}^2)^{1/2}$ where $\sigma_{i,\mathrm{w}}$ is the adopted uncertainty, $\sigma_i$ is the observational error associated with the $i$th radial velocity and $v_s\eta_i$ is the local radial velocity dispersion of the cluster according to the best-fit single-mass, isotropic King-Michie model (see §3). Of course, in order to convert $XYZ$ into the Galactic rest frame, we must correct for solar motion. To do this, we adopt a correction of

$$T_i = -108.1 \cos\alpha_i \cos\delta_i + 112.4 \sin\alpha_i \cos\delta_i - 172.1 \sin\delta_i. \tag{2}$$

to the radial velocity of an object at $\alpha_i, \delta_i$ (epoch 2000.0 coordinates). In deriving this correction we have adopted a basic solar motion of 16.5 km s$^{-1}$ toward $l = 53°, b = 25°$ (Binney and Tremaine 1987) and an LSR motion of 220 km s$^{-1}$ toward $l = 90°, b = 0°$ (Kerr and Lynden-Bell 1986). The best-fit space velocity for NGC 3201, corrected for solar motion and Galactic rotation, is then

$$v_{i,\mathrm{c}} = -302(\pm 25) \cos\alpha_i \cos\delta_i - 135(\pm 23) \sin\alpha_i \cos\delta_i - 168(\pm 23) \sin\delta_i \tag{3}$$

which implies a cluster velocity in the Galactic rest frame of $(\Pi, \Theta, \mathrm{Z}) = (-216 \pm 23, -214 \pm 24, 212 \pm 25)$. The magnitude of the velocity, $|\mathrm{V_s}| = 370 \pm 41$ km s$^{-1}$, is well below the local Galactic escape velocity of $\sim 475$ km s$^{-1}$ (Carney *et al.* 1988; Cudworth 1990). *We emphasize, however, that any space velocity derived in this fashion must be regarded as extremely uncertain since we have completely neglected rotation and tidal stripping, both of which are likely to play a role in explaining the large-scale trends in the velocity field.* Nevertheless, a proper motion study of NGC 3201 is clearly desirable since it would provide an direct test of our spectroscopically derived space velocity.

Regardless of the exact cause (or causes) of the observed structure in the velocity field, we have chosen to neglect it in modeling the cluster dynamics. This decision can be justified by assuming that the observed dependence of velocity on position agle is due entirely to rotation. The low ellipticity of NGC 3201 (like those of most other globular clusters, 95% of which have $\epsilon \leq 0.20$; White and Shawl 1987) suggests that ordered motions are dynamically unimportant. For instance, although neglecting a rotation of $\sim 1$ km s$^{-1}$ in the dynamical analysis will lead to *overestimates* of the cluster mass (Fischer *et al.* 1992a), the resulting errors will be at most a few percent (see §IVa of Pryor *et al.* 1986),

## 3. DYNAMICAL MODELS

In order to determine the form of the cluster mass function, the anisotropy of the VDP and several other interesting parameters including cluster mass, luminosity and M/L, we have fit single- and multi-mass King-Michie models to the observed *BV* SBPs and radial velocities. Since these models have seen widespread use in the study of globular clusters, the dynamical parameters derived from these models will be directly comparable to those of other clusters. In §3.2 we describe the results of modeling the observed SBPs and radial velocities using a nonparametric technique in which the form of the distribution function is not assumed *a priori* (Gebhardt and Fischer 1995).

### 3.1 *Single- and Multi-Mass Models*

Anisotropic, single-mass King-Michie models (King 1966; Michie 1963) assume a dis-

tribution function of the form

$$f(E,J) \propto e^{-\beta J^2}(e^{-E}-1) \tag{4}$$

where $E$ and $J$ refer to the energy and angular momentum of the cluster stars. Similarly, anisotropic, multi-mass models (Da Costa and Freeman 1976; Gunn and Griffin 1979) have, for each mass class, the distribution function

$$f_i(E,J) \propto e^{-\beta J^2}(e^{-A_i E}-1). \tag{5}$$

It is assumed that equipartition of energy in the cluster core has produced a dependence of the form $A_i \propto m_i$, where $m_i$ is the mean mass of the $i$th mass class. For each single-mass model, the anisotropy radius $r_a$ (the radius beyond which the velocity dispersion tensor is mostly radial) is held constant and the dimensionless central potential (King 1966), $W_0$, is varied until the best-fit values of the scale (or core) radius, $r_s$, the scale luminosity and the cluster concentration parameter, $c = \log(r_t/r_s)$, are obtained. In this way, we fit a grid of models with varying amounts of anisotropy to the cluster SBP. For each model, we determine the scale velocity, $v_s$, which gives the best match between the projected *model* VDP and the *observed* VDP (see § 3.1.2 for details of the fitting procedure). For the multi-mass models, we include another parameter, $x$, the global slope of the cluster mass function. Both $r_a$ and $x$ are then held constant for each model and the best-fitting scale radius, scale luminosity and concentration parameter are computed. The dimensionless, projected model VDP for the cluster giants is then scaled via maximum-likelihood to the measured velocities to yield $v_s$. The $BV$ SBPs for NGC 3201 are shown in Figure 6 along with the best-fit single-mass King-Michie models (with $r_a/r_s = \infty, 10, 5$ and 3). In the upper panel of Figure 7 we show the resulting VDP for NGC 3201; the solid line represents the best-fit, isotropic, single-mass model VDP, scaled by $v_s$ to the measured velocities. The LOWESS estimate of the velocity dispersion (see Gebhardt *et al.* 1994) used in the nonparametric modeling is indicated in the lower panel by the solid line. The velocity dispersion profile computed *in annular bins* is given in Table 5, whose columns record the bin number, sample size, radial range, median radius and intrinsic velocity dispersion estimated using the approach of Suntzeff *et al.* (1993) as well as that of Pryor and Meylan (1993). Both estimates of the binned dispersion profile are plotted in the lower pannel of Figure 7; the filled triangles indicate the Suntzeff *et al.* (1993) estimates of the dispersion while the filled squares represent those found using the Pryor and Meylan (1993) approach. For each bin, both the central location ("mean") and the scale ("dispersion") have been treated as free parameters.

A summary of the mass classes adopted for the multi-mass models is given in Table 6 which records, from left to right, the bin number, the lower bin boundary, the upper bin boundary and a description of the bin contents. We have followed the prescription of Pryor *et al.* (1989) in accounting for the evolved stars. Stars more massive for the main-sequence turnoff are assumed to have become cluster white dwarfs with objects having main-sequence masses in the ranges $8 - 4\mathrm{M}_{\odot}$, $4 - 1.5\mathrm{M}_{\odot}$ and $1.5 - 0.826\mathrm{M}_{\odot}$ (the mass at the tip of the red giant branch is taken to be $0.826\mathrm{M}_{\odot}$ after Bergbusch and VandenBerg 1992 and Brewer *et al.* 1993) assumed to have resulted in cluster white dwarfs with masses of 1.2, 0.7 and $0.5\mathrm{M}_{\odot}$, respectively. The neutron stars produced from higher mass stars are assumed to have been expelled from the cluster potential well, though the millisecond pulsars (Phinney 1992; Hut *et al.* 1992) and bright X-ray sources (Forman *et al.* 1978) seen in several clusters suggest that at least some of these objects contain neutron stars. Given the high space velocities observed for pulsars in the Galactic disk ($\simeq 210$ km s$^{-1}$; Lyne *et al.* 1982), it is unclear how neutron stars can remain bound to their respective clusters (*e.g.* the central escape velocity in NGC 3201 is $\lesssim 10$ km s$^{-1}$ according to our models). Moreover, neutron stars produced via Type II supernovae (which occurred approximately $10^{10}$ years ago in globular clusters) should have evolved to pulse periods in excess of $\simeq 1$ second (Bailyn 1993). Models in which millisecond pulsars are produced by the accretion-induced collapse of cluster white dwarfs (Michel 1987; Bailyn and Grindlay 1990) avoid these difficulties, so we have chosen to assume that neutron stars produced through Type II supernovae have been expelled from the cluster.

We have adopted a modified mass function (Pryor *et al.* 1989, 1991) of the form

$$\phi(\mathrm{M}) = \mathrm{M}^{-(1+x)} d\mathrm{M}, \qquad \text{for M} \geq 0.3\mathrm{M}_\odot \qquad (6)$$

$$\phi(\mathrm{M}) = \mathrm{M} d\mathrm{M}, \qquad \text{for M} < 0.3\mathrm{M}_\odot \qquad (7)$$

which is similar to that observed for local disk stars (Miller and Scalo 1979; Scalo 1986). The mass function is taken to have both high- and low-mass cutoffs, for which we adopt $\mathrm{M_H} = 8.0\mathrm{M}_\odot$ and $\mathrm{M_L} = 0.16\mathrm{M}_\odot$. As pointed out by Gunn and Griffin (1979), the choice of $\mathrm{M_L}$ is somewhat arbitrary — reducing the low-mass cutoff leads to models with enhanced numbers of low-mass stars at large radii and, consequently, to a higher inferred cluster masses. For stars fainter than the upper main sequence, we have used the isochrones of Bergbusch and VandenBerg (1992) to estimate the L/M of stars in the various mass bins (see below). Since their 16 Gyr, [Fe/H] = −1.26 and [O/Fe] = +0.55 isochrone ends at $0.1596\mathrm{M}_\odot$, we have chosen to truncate our mass function at $0.16\mathrm{M}_\odot$. In § 4 we discuss some of the consequences of adopting different low-mass cutoffs.

### 3.1.1 *Luminosity-to-Mass Ratios*

For each fitted model, we wish to compute two "Population" M/Ls — a *global* mass-to-light ratio (M/L) and a *central* mass-to-light ratio $(\mathrm{M/L})_0$. In order to derive the population M/Ls corresponding to our adopted mass function, we require a mean luminosity-to-mass ratio, L/M, for the component stars in each of our adopted mass bins. Some of the pitfalls involved in this rather uncertain process have been discussed by Pryor *et al.* (1986). Briefly, the mass bin containing the evolved cluster stars (red giants, subgiants, horizontal branch stars) and stars near the main-sequence turnoff contributes virtually all of the cluster light; cluster population M/Ls therefore depend sensitively on the L/M adopted for this bin. A *photometrically and spatially complete* luminosity function for these stars is therefore required since, for NGC 3201, the $\mathrm{L}_V/\mathrm{M}$ of stars in this bin varies from $\simeq 840$ at the tip of the red giant branch to $\simeq 2$ at the main-sequence turnoff (Bergbusch and VandenBerg 1992).

We have therefore used our wide-field $BV$ CCD images to perform star counts in NGC 3201 of stars brighter than $V = 19.97$ (corresponding to a mass of $0.75\mathrm{M}_\odot$). Our counts are photometrically complete for stars of this brightness (the majority of which are expected to fall within the CCD fields shown in Figure 1). Each star brighter than this limiting magnitude was assigned a position (based on its location in the cluster color-magnitude diagram) on either the 16 Gyr, [Fe/H] = –1.26, [O/Fe] = +0.55 isochrone of Bergbusch and VandenBerg (1992) or the corresponding horizontal branch evolutionary sequence of Dorman (1992); probable field stars were rejected from the analysis. Both the luminosity and mass of the individual stars were added to derive a mean L/M for the stars in the bin — based on counts of 7660 upper main-sequence, subgiant, red giant branch stars and 237 horizontal branch stars, we adopted mean L/Ms of $\mathrm{L}_V/\mathrm{M} = 10.03$ and $\mathrm{L}_B/\mathrm{M} = 9.33$ for stars in the range $0.75 - 0.826\mathrm{M}_\odot$. (Throughout this paper, we give mass-to-light ratios in solar units.) Since our photometry is not deep enough to derive a reliable luminosity function for the fainter main sequence stars, we used the same Bergbusch and VandenBerg (1992) isochrone to derive a mean L/M for each of the remaining bins. Although these L/Ms vary with the adopted mass function slope, the dependence is very weak, amounting to a $\sim 3\%$ decrease in the mean L/M of the main-sequence stars in the bin as $x$ increases from 0.0 to 2.0.

### 3.1.2 *Fitting the Models*

For each model, we have minimized

$$\chi^2 = \sum_{j=1}^{\mathrm{N}} \frac{1}{\Delta_j{}^2}[\mathrm{L}_s\sigma(r_j/r_s) - \mu_j]^2 \tag{8}$$

in order to get the best-fit scale radius, $r_s$, and scale luminosity, $\mathrm{L}_s$. Here $\mu_j$ are the measured surface brightnesses and $\sigma(r_j/r_s)$ are the projected model surface brightnesses at radii $r_j$. The corresponding uncertainties in $\mu_j$ are given by $\Delta_j$. The $\chi^2$ goodness-of-fit statistic is computed for each fitted model and the reduced gravitational potential, $W_0$, is varied until the computed $\chi^2$ is minimized. The maximum-likelihood estimators for the scale velocity, $v_s$, and the cluster systemic velocity, $v_0$, are then found by solving (Gunn and Griffin 1979)

$$\sum_{i=1}^{N} \frac{v_i}{(v_s^2\eta_i^2 + \sigma_i^2)} - v_0 \sum_{i=1}^{N} \frac{1}{(v_s^2\eta_i^2 + \sigma_i^2)} = 0 \tag{9}$$

$$\sum_{i=1}^{N} \frac{(v_i - v_0)^2}{(v_s^2\eta_i^2 + \sigma_i^2)} - \sum_{i=1}^{N} \frac{1}{(v_s^2\eta_i^2 + \sigma_i^2)} = 0 \tag{10}$$

where $v_i$ and $\sigma_i$ are the measured radial velocities and corresponding uncertainties; $\eta_i$ refers to the projected, dimensionless (model) velocity dispersion at the radius corresponding to $v_i$. According to our models, $v_0$ ranges from 494.1 – 494.3 km s$^{-1}$.

We then compute two "Dynamical" M/Ls (once again, a *global* and a *central* M/L) using our fitted model and the maximum-likelihood estimator for the scale velocity. In order for the model to be considered acceptable, both dynamical M/Ls should match the population M/Ls computed with the assumed mass function (of course, the best models should also have relatively low $\chi^2$ values). For each model, we then compute a number of cluster parameters which are summarized in Tables 7 and 9 (for the $V$-band) and Tables 8 and 10 (for the $B$-band). Fitted cluster parameters are given in Tables 7 and 8 which record, from left to right, the anisotropy radius in units of the scale radius, the mass function slope, the cluster concentration parameter, the dimensionless central potential $W_0$, the scale radius in pc, the central surface brightness $\mu_0$ in $L_\odot$ pc$^{-2}$, the reduced $\chi^2$ for the fit to the SBP, the probability of meeting or exceeding this $\chi^2$, the scale velocity in km s$^{-1}$, the central and global population M/Ls and the central and global dynamical M/Ls. Derived cluster parameters are recorded in Tables 9 and 10 whose columns contain, from left to right, the anisotropy radius in units of the scale radius, the mass function slope, the scale radius in pc, the half-mass radius $r_h$ in pc, the tidal radius $r_t$ in pc, the model central velocity dispersion $v_s\eta_0$ in km s$^{-1}$, the total cluster luminosity L in $L_\odot$, the central luminosity density $\Sigma_0$ in $L_\odot$ pc$^{-3}$, the total cluster mass M in $M_\odot$, the central mass density $\rho_0$, the mean density inside the half-mass radius $\rho_h$, the mean density inside the tidal radius $\rho_t$ (all in $M_\odot$ pc$^{-3}$), the logarithm of the half-mass relaxation time $t_{r0}$ in years (Lightman and Shapiro 1978) and the logarithm of the half-mass relaxation time $t_{rh}$ in years (Spitzer and Hart 1971).[7]

### 3.1.3 *Monte Carlo Simulations*

Also recorded in Tables 7 – 10 are the one sigma uncertainties for each of the above parameters, determined through Monte Carlo experiments like those described by Pryor *et al.* (1989) and Fischer *et al.* (1992b). Briefly, 1000 datasets were generated from the best-fit model using the estimated uncertainties in the actual SBP. Both the artificial SBPs and the simulated radial velocities have points at the identical distance (and, for the velocities, identical position angle) as the actual data. For the radial velocity simulations, we have followed the prescription of Fischer *et al.* (1992b) and have generated random three-dimensional positions as well as radial and tangential velocities for each of the measured stars. The velocities were then projected onto the plane of the sky and a random measurement error (based on the actual uncertainty) included. Based on model fits to these 1000 simulated datasets, the rms dispersion about the mean of each parameter has been taken as the one sigma uncertainty.

Of course, the uncertainty derived in this manner represents only the internal error in the fitted parameter. The large spread in the best-fit parameters computed from the various models (for example, the cluster mass ranges from $1.1\times10^5$ $M_\odot$ to $5.4\times10^5$ $M_\odot$ based on fits to the $V$-band SBP) demonstrates that the true errors are likely to be much larger. For example, it is now recognized that vastly different density profiles are capable of providing equally impressive fits to the SBPs of most globular clusters (*e.g.* Merritt 1993), so that cluster parameters derived using the King-Michie formalism need not reflect the true physical state of the cluster. To investigate this possibility, we have modeled the observed SBPs and radial velocities using nonparametric techniques which make no *a priori* assumption about the cluster distribution function.

[7] For the single-mass models, $t_{r0}$ and $t_{rh}$ have been computed using a stellar mass of $0.65M_\odot$.

### 3.2 *Nonparametric Models*

Since a complete discussion of the nonparametric technique may be found in Gebhardt and Fischer (1995), only a brief description is given here. In all cases, we have assumed that the stellar velocities are isotropic — the extension to anisotropic velocities will be reported in the near future (Gebhardt and Merritt 1995). The technique requires both a cluster SBP and VDP (the latter is estimated using a LOWESS fit to the data; see Figure 7). We then estimate the deprojected quantities through the Abel integrals

$$\Sigma(r) = -\frac{1}{\pi}\int_r^{r_t} \frac{d\mu}{dR}\frac{dR}{\sqrt{R^2 - r^2}} \tag{11}$$

$$\Sigma(r)v_r^2(r) = -\frac{1}{\pi}\int_r^{r_t} \frac{d(\mu\sigma_p^2)}{dR}\frac{dR}{\sqrt{R^2 - r^2}} \tag{12}$$

where $\Sigma(r)$ is the luminosity density, $\mu$ is the surface brightness, and $\sigma_p$ and $v_r$ are the projected and deprojected velocity dispersions, respectively. In practice, the above integrals cannot be evaluated out to the tidal radius since the cluster surface brightness and velocity dispersion near the tidal radius are poorly known. For this reason, the point where the velocity dispersion is last measured has been taken as the upper limit (although we do not consider the contributions from beyond this point, the effect is non-negligible only near the tidal radius). Once the deprojected quantities are in hand, we can use the Jeans equation to estimate the mass and mass density (Binney and Tremaine 1987):

$$M(r) = -\frac{rv_r^2}{G}\left(\frac{d\,\ln\Sigma}{d\,\ln r} + \frac{d\,\ln v_r^2}{d\,\ln r}\right) \tag{13}$$

and

$$\rho(r) = \frac{1}{4\pi r^2}\frac{dM}{dr}\ . \tag{14}$$

Since these equations involve two and one half derivatives of both the surface brightness and the projected velocity dispersion, a certain amount of smoothing of the (noisy) data is required. We use a spline smoother with the smoothing parameter chosen by generalized cross validation (Wahba 1990). All calculations are performed in logarithmic space to avoid enhanced weighting of the higher values when using the spline fitter.

We then compute the cluster mass density and M/L profiles. Observational biases and confidence bands are determined through Monte Carlo simulations in which artificial datasets are generated by randomly choosing a velocity from a Gaussian distribution with the standard deviation given by the dispersion profile at the radius of each observation and the uncertainty of each velocity measurement. The procedure described above is then used to compute, in a completely analogous fashion, the mass density and M/L profiles for each simulation. By generating 1000 simulations, we have a distribution in mass density and M/L at each point in our profile which we use to measure both the mode and the 95% confidence band. The central location of the simulation distribution minus the initial estimate of the mass density is then adopted as the estimate for the bias. Once determined, we must correct for the bias by adding it back into the original estimate. The confidence bands are correspondingly shifted for the bias as well, though the confidence bands require *twice* the bias to be added since the simulations have a bias from both the technique and from the original estimate. We have assumed that the velocity distribution at each radius is Gaussian. The tidal cutoff ensures that this is *not* the case, and a fully nonparametric technique would need to include a proper estimation of the velocity distribution at each radius. Nevertheless, we feel that deviations from the assumed Gaussian distribution are likely to be small enough to have negligible effect on our results.

Figure 8 shows the mass density profile and M/L profile of NGC 3201 computed in this fashion. The solid lines are the bias-corrected mass density and M/L estimates while the dotted lines indicate the 95% confidence bands. Although the VDP and SBP extend to both smaller and larger radii than are plotted, the confidence bands become so large that the estimates of mass density are essentially meaningless in these regions. The dashed lines in Figure 8 indicate the cluster mass density and M/L profiles according to one of the best-fit, multi-mass King-Michie models ($r_a/r_s = \infty$, $x = 1.0$) from §3.1. In general, the profiles determined using the different approaches show very good agreement — the mass density and M/L profiles determined with the the multi-mass models fall within the 95% confidence bands of the nonparametric profiles for virtually all radii. The M/L profile determined via the King-Michie approach shows a systematic rise in the outer regions of the cluster (a consequence of the assumption of energy equipartition among the various mass species), whereas that derived from the nonparametric models shows a rather low central value of $M/L_V \approx 1$ and a steady rise to $M/L_V \approx 4$ at 10 pc.

## 4. RESULTS

### 4.1 *Previous Work on NGC 3201*

How do the results of our modeling compare to previous studies? The first attempt to measure the cluster SBP was that of King *et al.* (1968) who performed star counts on photographic plates, tracing the SBP out to a radius of approximately 20′. Their best-fit isotropic, single-mass model was found to have $c = 1.56$. The lone previous dynamical study of NGC 3201, that of Da Costa *et al.* (1993), combined the King *et al.* (1968) star counts with more recent photoelectric aperture photometry and CCD surface photometry to derive a somewhat lower concentration of $c = 1.38$, suggesting that King *et al.* (1968) underestimated the cluster background (since NGC 3201 is a low latitude cluster, background contamination is rather severe). NGC 3201 was also included in the CCD survey of globular cluster structural parameters of Trager *et al.* (1995), who found $c = 1.31$ and $r_c = 1.45'$, in excellent agreement with our values of $c = 1.26$ and $r_c = 1.46'$ ($V$-band SBP) and $c = 1.33$ and $r_c = 1.38'$ ($B$-band SBP).

Da Costa *et al.* (1993) combined their SBP with mean radial velocities for 92 cluster giants (included in the present sample and published in Côté *et al.* 1994) to derive a cluster M/L of 1.6±0.5 using isotropic, single-mass King-Michie models. (For comparison, they found $v_0 = 493.0 \pm 1.0$ km s$^{-1}$ and $\sigma_0 = 4.4 \pm 0.5$ km s$^{-1}$.) This is in good agreement with our values of $M/L_V = 1.62 \pm 0.11$ and $M/L_B = 1.66 \pm 0.11$ (also for isotropic, single-mass models). However, Da Costa *et al.* (1993) assumed a reddening of $E(B-V) = 0.28$ and a cluster distance of 4.5 kpc, compared to our values of $E(B-V) = 0.21$ and R = 5.1 kpc. Scaling to our distance and reddening reduces their M/L to 1.4±0.5 — still in good agreement with our findings.

### 4.2 *Mass-to-Light Ratios*

Using spectra of the integrated cluster light, Illingworth (1976) found the M/Ls of ten centrally concentrated globular clusters to fall in the range 0.9 to 2.6 with a mean of 1.6, in excellent agreement with the NGC 3201 mass-to-light ratio computed from single-mass King models. For a sample of 32 clusters with reliable central velocity dispersions, Mandushev *et al.* (1991) derived a mean M/L of $\simeq$ 1.2 using single-mass King models, whereas Pryor and Meylan (1993) compiled velocity dispersion data for 56 galactic globular clusters and modeled the available SBPs and velocities with multi-mass King models, reporting a mean of 1.7±0.9 for the entire sample. (Since this value is sensitive to outliers, they also report a mean of 2.3±1.1 based on biweight estimators). Imposing the criterion that an acceptable model must fit have similar (1) *central* population and dynamical M/Ls and (2) *global* population and dynamical M/Ls, we find good agreement (for both the $B$ and $V$ SBPs) for models having global mass-to-light ratios in the range 1.8 – 2.5. The best-fitting models (which have $x = 0.75$; see below) have $\mathrm{M/L}_V \simeq \mathrm{M/L}_B = 2.0\pm0.2$ and $(\mathrm{M/L}_V)_0 \simeq (\mathrm{M/L}_B)_0 = 1.75\pm0.09$. It therefore appears that the M/L of NGC 3201 does not differ significantly from mean of the Galactic globular cluster population.

### 4.3 *Mass Function Slope*

Inspection of Tables 7 - 10 shows that models with mass function slopes in the range $0.5 \lesssim x \lesssim 1.0$ provide the best match between the central and global dynamical and population M/Ls. A grid of models with stepsize $\Delta x = 0.1$ fit over this interval showed best agreement for slopes of $x = 0.7$ and 0.8 (for the respective $V$- and $B$-band SBPs); we therefore adopt a best-fit value of $x = 0.75 \pm 0.25$. How sensitive is this value to the form of the adopted mass function? For mass functions with no transition at $\mathrm{M} = 0.3\mathrm{M}_\odot$ (*i.e.* single exponent mass functions), models with $x = 0.4 - 0.5$ show the best agreement between the population and dynamical M/Ls. Similarly, experiments with a variety lower mass cutoffs confirmed the findings of earlier workers (*e.g.* Gunn and Griffin 1979) who noted that changing $M_L$ has relatively little effect on the observable properties of the models — for instance, reducing $M_L$ from $0.16M_\odot$ to $0.05M_\odot$ lowers our best-fit value of $x = 0.75 \pm 0.25$ by $\approx 0.25$. In both cases, models having $x \gtrsim 1.0$ yield dynamical M/Ls which are unacceptably large compared to those implied by the adopted mass function.

NGC 3201 has recently been the subject of a photometric study by Brewer *et al.* (1993) who used $BVI$ CCD images for a field located approximately seven core radii from the cluster center to study the cluster luminosity and mass functions. Only for the $I$-band did their data extend faint enough to reliably estimate the mass function exponent. For stars in the approximate range 0.40 – $0.22M_\odot$, their mass function was found to rise sharply with index $x = 2.0 \pm 0.3$ (though they point out that if their lowest mass data point is excluded, this estimate drops to $1.5\pm0.4$). In either case, their measured mass function exponent is somewhat larger than our dynamically determined value of $x = 0.75 \pm 0.25$. Given the differences in the respective forms of the adopted mass functions (and the fact that our best-fit value applies to a considerably different mass range: 8.0–$0.3M_\odot$) such a discrepancy is perhaps not suprising.

Since possible correlations of mass function slope with other cluster parameters (Piotto 1991; Richer *et al.* 1991; Capaccioli *et al.* 1993) have important implications for the formation and dynamical evolution of not only the Galactic globular cluster system but also the Galactic halo, it is natural to ask whether or not this slope agrees with previously suggested trends. The left panel of Figure 9 shows the global mass function slope plotted against the distance from the Galactic plane, $|Z_G|$, for the 17 clusters studied by Capaccioli *et al.* (1993) with two new additions: NGC 362 (Fischer *et al.* 1993) and NGC 3201 (open square). Global mass function slope versus Galactocentric distance, $|R_G|$, is given in the second panel of Figure 9 for the same sample of 19 clusters (all $|Z_G|$ and $|R_G|$ are taken from the compilation of Djorgovski 1993 who assumes a solar Galactocentric distance of 8 kpc, as opposed to Capaccioli *et al.* 1993 who adopt 8.8 kpc following Harris 1976). In both cases, NGC 3201 agrees with the previously identified trends (which Capaccioli *et al.* 1993 interpreted as evidence for the dynamical evolution of a universal globular cluster mass function due to Galactic disk-shocking).

Figure 10 shows the global mass function slope for the same cluster sample plotted against the logarithm of the half-mass relaxation time, $t_{rh}$, taken from Djorgovski (1993) for all clusters except NGC 3201 (for which we adopt log $t_{rh} = 8.775$; see Tables 8 and 10) and the cluster disruption time, $t_d$, defined by Richer *et al.* (1991) as the inverse of the cluster destruction rate calculated by Aguilar *et al.* (1988). Richer *et al.* (1991) found a correlation between $t_d$ and the mass function slope below $0.4M_\odot$ based on a sample of six clusters with deep luminosity functions. However, with NGC 362 and NGC 3201 added to the Capaccioli *et al.* (1993) sample, no such correlation between $x$ and $t_d$ (or $t_{rh}$) is evident in Figure 10, as previously noted by Capaccioli *et al.* 1993.

### 4.4 *Anisotropy*

Unlike most previous studies on cluster dynamics which have been able to place only rather weak limits on the cluster M/L and mass function slope, our large sample of radial velocities has allowed us to put more stringent constraints on these parameters. However, NGC 3201 was targeted for study principally on the basis of its high systemic radial velocity and without consideration to its overall luminosity — since it is an intrinsically sparse cluster with a SBP extending over only $\simeq$ three orders of magnitude in luminosity (compared to the five or more decades available for some clusters such as M15, Newell and O'Neill 1978 and M3, Da Costa and Freeman 1976) the SBPs provide relatively little information on the anisotropy of the stellar velocities.

As previously mentioned, mass function slopes in the range $0.5 \lesssim x \lesssim 1.0$ provide the best fits to the observed SBPs and VDP. In general, for values of $x$ in this range, only those models with very strong anisotropy ($r_a/r_s = 3$) are ruled out by the observed

SBP. Although isotropic models usually provide the best match to the data, models having anisotropy radii in the range $5 \lesssim r_a/r_s \lesssim 10$ often provide acceptable fits. (Models with $r_a/r_s = 5$ are weakly ruled out if $M_L$ is reduced from $0.16 M_\odot$ to $0.05 M_\odot$.) While definite conclusions about the velocity anisotropy in NGC 3201 would therefore be premature, it appears that both isotropic and weakly anisotropic orbits provide the most impressive fits to the cluster SBPs. Because of the low cluster velocity dispersion, the radial velocities provides no further discrimination[8] — the agreement between the observed and theoretical VDPs is excellent (see Figure 7) for each of the $r_a/r_s = \infty, 10$ and 5 models (the fit to the $r_a/r_s = 3$ models is marginally inferior). The excellent match of the observed VDP to the isotropic and weakly anisotropic models may be a consequence of two-body relaxation in the core and the influence of the Galactic tidal field near the cluster boundary — Oh and Lin (1992) recently constructed Fokker-Planck/three-body integration models for the tidal regions of globular clusters and found that the Galactic tidal torque induces isotropy near the limiting radius, consistent with our findings for NGC 3201.

## 5. SUMMARY

We have carried out a dynamical analysis of the nearby globular cluster NGC 3201 based on *B*- and *V*-band CCD measures of the cluster surface brightness profile and 857 radial velocities for 399 cluster giants. The observed VDP extends over the full range in cluster radius with member giants detected as far as $32'$ ($\sim r_t$) from the cluster core. The median difference in radial velocity for stars on either side of an imaginary axis stepped through the cluster in $1^\circ$ increments shows a maximum amplitude of $1.22 \pm 0.25$ km s$^{-1}$. Monte Carlo experiments suggest that this observed amplitude is significant at the 99.2 % level. Possible explanations of this observation include: (1) cluster rotation (supported by the good agreement between the observed cluster ellipticity and photometric minor axis orientation and that expected for a rotationally-flattened oblate spheroid); (2) preferential stripping of stars on prograde orbits near the limiting radius; (3) the projection of the cluster space velocity onto the plane of the sky and; (4) a slight drift in the Argus velocity zero point. It is difficult to identify which of these processes is the dominat one and we suspect that each may play a role in explaining the observed structure in the velocity field.

[8] Of course, high-quality proper motions for the cluster members studied here would provide a better diagnostic of the anisotropy of the VDP. Unfortunately, proper motions of the requisite precision do not yet exist for NGC 3201.

Single-mass model fits to the observed SBPs show good agreement with those of Da Costa *et al.* (1993) and Trager *et al.* (1995). The cluster M/Ls derived from these single component models ($M/L_V \simeq M/L_B \simeq 1.7 \pm 0.1$) are also in good agreement with that found in the only previous dynamical study of NGC 3201 based on radial velocities of individual member stars (Da Costa *et al.* 1993). Multi-mass and nonparametric models yield slightly higher values, $M/L_V \simeq M/L_B \simeq 2.0 \pm 0.2$, and both approaches suggest the cluster M/L increases monotonically in the range 1.5 – 10 pc. The best-fit, multi-mass models have mass function slopes of $x \simeq 0.75 \pm 0.25$, consistent with the $x(|Z_G|)$ and $x(R_G)$ correlations observed by Capaccioli *et al.* (1993). Due to the low cluster luminosity, we are able to place only weak constraints on the anisotropy of VDP — isotropic orbits generally provide the best fits to the observations though models with anisotropy radii as small as $r_a/r_s = 5$ are still capable of providing impressive fits.

An obvious extension of the present work is the measurement of proper motions for the radial velocity members observed in this study. Unfortunately, the requisite observations are exceedingly difficult since the centermost stars in NGC 3201 are expected to show proper motions of only $\approx$ 20 milliarcseconds per century. Such observations would not only yield the two components of the VDP needed to solve the non-rotating Jeans equation but would also obviate the assumption of an isotropic VDP implicit in our nonparametric models.

The authors thank the TACs of both the Observatories of the Carnegie Institute of Washington and the Cerro Tololo Inter-American Observatory for the allocation of telescope time. Thanks also to Mike Irwin for providing the APM/UK-Schmidt plate scans used in this study, and to Ruth Peterson for several useful comments. This research was funded in part by the Natural Sciences and Engineering Council of Canada. The radial velocities and surface brightness profiles used in this analysis are available in machine-readable form — contact the first author for details.

## Figure Captions

**Figure 1** – The location with respect to the cluster center (indicated by the dot) of our five innermost CCD fields. The circle represents the NGC 3201 core radius ($r_s = 1.46'$) determined from single-mass, isotropic King-Michie models fits to the $V$-band SBP. Each CCD frame measures $10.4'$ on a side with east to the top and north to the left. A series of partly overlapping fields (extending out to about $50'$ from the NGC 3201 center in both of the north and south directions) have been omitted for clarity.

**Figure 2** – (Upper Panel) Heliocentric radial velocity versus distance from cluster center for the same sample of cluster members. The dashed line at 494.2 km s$^{-1}$ indicates the mean cluster velocity according to the maximum-likelihood technique of Gunn and Griffin (1979). (Middle Panel) Heliocentric radial velocity versus position angle $\Phi$ for 399 NGC 3201 members. (Lower Panel) Annular bins of median radial velocity versus position angle $\Phi$ for the same sample of 399 stars (see Table 3).

**Figure 3** – (Upper Panel) The difference in median radial velocity for stars on either side of an axis at position angle $\Phi$. Also shown is the best-fit sine curve which indicates a position angle of $-7 \pm 12°$ for the cluster rotation axis. (Lower Panel) Histogram of amplitudes of the best-fit sine curves for 1000 Monte Carlo simulations of the data with no rotation. The observed amplitude of $1.22 \pm 0.25$ km s$^{-1}$ (indicated by the arrow) is significant at the 99.2% level.

**Figure 4** – The $V_{\rm rot}/\sigma$ versus ellipticity ($\epsilon = 1 - b/a$) for a rotationally-flattened, oblate spheroid (dashed line) with an isotropic velocity dispersion tensor. NGC 3201 is indicated by the filled circle. Projection effects tend to move points on the curve in the approximate direction of the origin.

**Figure 5** – The location of the 200 outermost NGC 3201 members on the plane of the sky. Objects indicated by open circles have $(v - v_0) < 0$ while those shown

as open squares have $(v - v_0) > 0$ (where $v_0 = 494.2$ km s$^{-1}$). In all cases, the size of the point is proportional to the magnitude of velocity residual. A cluster tidal radius of $26.8'$ ($V$-band, single-mass models) is indicated by the dotted circle, though this parameter is very poorly constrained by the observations.

**Figure 6** – (Upper Panel) The $V$-band surface brightness profile for NGC 3201. Circles indicate the results of CCD surface photometry while the square indicate points determined by CCD star counts. The best-fitting single-mass, King-Michie models are also shown as the solid (isotropic), dotted ($r_a/r_s = 10$), short-dashed ($r_a/r_s = 5$) and long-dashed ($r_a/r_s = 3$) lines. (Lower Panel) Same as above except for the $B$-band.

**Figure 7** – (Upper Panel) The velocity dispersion profile for NGC 3201. The solid line represents the profile expected on the basis of the best-fit, $V$-band, isotropic, single-mass model. Each point represents the absolute difference between the stellar velocity and the fitted mean cluster velocity for our 399 program objects. (Lower Panel) The LOWESS estimate of the velocity dispersion (solid line) and the corresponding 90% confidence bands (dotted lines). For comparison, we also show the binned velocity dispersion profile derived using the Suntzeff *et al.* (1993) variant of the Armandroff and Da Costa (1986) technique (filled triangles) as well as that found using the Pryor and Meylan (1993) maxmimum-likelihood estimators (filled squares). The dispersion in each bin has been computed using the associated average bin velocity, rather than mean velocity of the whole sample. The outermost bin contains 39 stars; all others contain 40. For ease of comparison, the King-Michie profile shown in the upper panel is indicated by the dashed line.

**Figure 8** – (Upper Panel) The mass density profile of NGC 3201 according to one of the best-fit multi-mass, King-Michie models ($r_a/r_s = \infty$, $x = 1.0$; dashed

line). The variation in mass density determined from nonparametric modeling of the individual velocities is given by the solid line (the dotted lines indicate 95% confidence bands). (Lower Panel) The variation in M/L$_V$ computed with the same King-Michie model (dashed line). The same profile determined with nonparametric models is shown as the solid line; the dotted lines represent 95% confidence bands.

**Figure 9** – (Left Panel) Global mass function slope, $x$, versus distance from the Galactic disk, $|Z_G|$, for the 17 clusters (filled circles) of Capaccioli *et al.* (1993) and NGC 362 (Fischer *et al.* 1993). NGC 3201 is indicated by the open square. (Right Panel) Global mass function slope versus distance from the Galactic center, $R_G$, for the same 19 clusters. Once again, NGC 3201 has been included as the open square. In all cases, we have taken $|Z_G|$ and $R_G$ from the compilation of Djorgovski (1993).

**Figure 10** – (Left Panel) Global mass function slope, $x$, versus the logarithm of the half-mass relaxation timescale for the 17 clusters (filled circles) of Capaccioli *et al.* (1993), NGC 362 (Fischer *et al.* 1993) and NGC 3201. Relaxation timescales for all clusters except NGC 3201 are taken from Djorgovski (1993). For NGC 3201 (open square), we have adopted $\log t_{rh} = 8.775$, the average of the determinations based on single-mass, isotropic King-Model model fits to the $B$- and $V$-band surface brightness profiles, in excellent agreement with the value of 8.79 found by Djorgovski (1993). (Right Panel) Global mass function slope versus the "disruption timescale" according to Aguilar *et al.* (1988) for NGC 362, NGC 3201 and 16 of the clusters (filled circles) in the Capaccioli *et al.* (1993) sample (NGC 5053 was omitted by Aguilar *et al.* 1988 in their study of Galactic globular cluster system). NGC 3201 is shown as the open square.

**Mailing Addresses:**

Patrick Côté

Dominion Astrophysical Observatory, Herzberg Institute of Astrophysics

National Research Council, 5071 West Saanich Road, Victoria, BC, V8X 4M6, Canada

Philippe Fischer

AT&T Bell Laboratories, 600 Mountain Ave., 1D-316

Murray Hill, NJ 07974, USA

Karl Gebhardt

Department of Astronomy, University of Michigan

Dennison Building, Ann Arbor, MI 48109-1090, USA

Douglas L. Welch

Department of Physics and Astronomy, McMaster University

Hamilton, ON, L8S 4M1 Canada

TABLE 1

General Cluster Parameters

| Parameter | NGC 3201 | Source |
|---|---|---|
| $\alpha$ (2000.0) | $10^{h}\ 17^{m}\ 36.75^{s}$ | Shawl and White (1986) |
| $\delta$ (2000.0) | $-46^{\circ}\ 24'\ 40.2''$ | Shawl and White (1986) |
| $l^{II}$ | $277^{\circ}\ 13'\ 40.8''$ | Shawl and White (1986) |
| $b^{II}$ | $8^{\circ}\ 38''\ 29.0''$ | Shawl and White (1986) |
| $E(B-V)$ | 0.21±0.02 mag | Lee (1977) |
| $(m-M)_V$ | 14.20±0.15 mag | Brewer *et al.* (1993) |
| $R_{\odot}$ | 5.1±0.4 kpc | Brewer *et al.* (1993) |
| Age | 15±2 Gyr | Brewer *et al.* (1993) |
| [Fe/H] | −1.3±0.1 | Brewer *et al.* (1993) |
| $\overline{b/a}$ | 0.88±0.01 | White and Shawl (1987) |
| $v_0$ [a] | 494.0±0.2 km $s^{-1}$ | This paper |
| $\sigma_{\rm int}$ [b] | 3.7±0.1 km $s^{-1}$ | This paper |
| $\sigma_{\rm 0,int}$ [b] | 3.9±0.3 km $s^{-1}$ | This paper |

a – The systemic cluster velocity according to the estimator of Suntzeff *et al.* (1993) using the entire sample of 399 cluster members (i.e. candidate binaries excluded). The quoted error in the mean velocity refers to the *internal* uncertainty and neglects the zero point uncertainty of about 1.0 km $s^{-1}$.

b – Intrinsic, one-dimensional velocity dispersions according to the Suntzeff *et al.* (1993) estimator; $\sigma_{\rm int}$ has been computed using the entire sample of 399 cluster members, whereas $\sigma_{\rm 0,int}$ refers to the dispersion obtained using only those 93 stars within one core radius of the cluster center.

TABLE 2

$BV$ Surface Brightness Profiles

| R (pc) | $\mu_V$ ($L_{V\odot}$ pc$^{-2}$) | Type[a] | R (pc) | $\mu_B$ ($L_{B\odot}$ pc$^{-2}$) | Type[a] |
|---|---|---|---|---|---|
| 0.25 | 2361.8±1590.6 | SP | 0.25 | 2566.0±1276.4 | SP |
| 0.59 | 1604.2± 414.3 | SP | 0.59 | 1876.1± 371.9 | SP |
| 0.96 | 1928.6± 479.3 | SP | 0.96 | 2076.1± 391.8 | SP |
| 1.33 | 1517.9± 401.3 | SP | 1.33 | 1634.0± 349.0 | SP |
| 1.71 | 1002.9± 164.6 | SP | 1.71 | 1162.1± 151.7 | SP |
| 2.09 | 906.2± 162.4 | SP | 2.09 | 975.5± 160.1 | SP |
| 2.46 | 609.9± 139.7 | SP | 2.46 | 681.2± 98.9 | SP |
| 2.84 | 566.6± 98.4 | SP | 2.84 | 668.7± 116.1 | SP |
| 3.20 | 454.2± 20.9 | SC | 3.22 | 516.7± 72.3 | SP |
| 3.22 | 462.6± 94.4 | SP | 3.22 | 512.0± 26.1 | SC |
| 3.59 | 410.4± 18.9 | SC | 3.60 | 450.0± 24.9 | SC |
| 3.60 | 395.6± 68.7 | SP | 3.60 | 475.6± 75.5 | SP |
| 3.97 | 347.7± 16.0 | SC | 3.97 | 384.8± 21.3 | SC |
| 3.98 | 341.4± 88.2 | SP | 3.98 | 381.3± 72.0 | SP |
| 4.35 | 254.2± 27.1 | SP | 4.34 | 323.0± 17.9 | SC |
| 4.35 | 308.5± 15.7 | SC | 4.35 | 291.9± 33.8 | SP |
| 4.73 | 263.8± 13.4 | SC | 4.72 | 302.9± 16.8 | SC |
| 4.73 | 273.7± 58.5 | SP | 4.73 | 326.0± 56.6 | SP |
| 5.10 | 219.4± 12.1 | SC | 5.10 | 238.4± 15.4 | SC |
| 5.11 | 171.1± 33.0 | SP | 5.11 | 196.4± 38.9 | SP |
| 5.47 | 180.8± 10.9 | SC | 5.48 | 182.5± 12.7 | SC |
| 5.49 | 238.4± 121.2 | SP | 5.49 | 236.2± 113.8 | SP |
| 5.87 | 175.9± 43.7 | SP | 5.87 | 187.6± 43.0 | SP |
| 5.87 | 154.6± 10.0 | SC | 5.87 | 160.4± 11.1 | SC |
| 6.24 | 111.0± 5.1 | SP | 6.24 | 171.1± 11.0 | SC |

TABLE 2 cont'd

*BV* Surface Brightness Profiles

| R (pc) | $\mu_V$ ($L_{V\odot}$ pc$^{-2}$) | Type[a] | R (pc) | $\mu_B$ ($L_{B\odot}$ pc$^{-2}$) | Type[a] |
|---|---|---|---|---|---|
| 6.25 | 151.8± 9.1 | SC | 6.24 | 131.0± 6.0 | SP |
| 6.62 | 156.1± 34.8 | SP | 6.62 | 184.2± 36.5 | SP |
| 6.62 | 133.4± 8.6 | SC | 6.63 | 133.4± 10.5 | SC |
| 6.99 | 112.0± 7.8 | SC | 7.00 | 131.0± 9.7 | SC |
| 7.00 | 124.0± 18.5 | SP | 7.00 | 145.0± 16.8 | SP |
| 7.37 | 96.7± 7.1 | SC | 7.38 | 107.0± 8.9 | SC |
| 7.38 | 88.2± 8.6 | SP | 7.38 | 100.3± 9.8 | SP |
| 7.75 | 72.7± 21.1 | SP | 7.75 | 85.8± 34.5 | SP |
| 7.76 | 83.4± 6.9 | SC | 7.76 | 101.2± 8.9 | SC |
| 8.12 | 81.1± 15.6 | SP | 8.12 | 90.6± 17.4 | SP |
| 8.13 | 72.0± 7.0 | SC | 8.13 | 79.7± 8.5 | SC |
| 8.49 | 65.7± 6.4 | SC | 8.51 | 86.5± 37.9 | SP |
| 8.51 | 89.0± 49.4 | SP | 8.51 | 62.7± 7.6 | SC |
| 8.88 | 66.9± 27.4 | SP | 8.88 | 64.5± 28.7 | SP |
| 8.90 | 63.3± 6.2 | SC | 8.90 | 68.1± 7.6 | SC |
| 9.27 | 53.6± 5.7 | SC | 9.26 | 59.3± 7.2 | SC |
| 9.66 | 43.0± 5.2 | SC | 9.65 | 55.6± 6.7 | SC |
| 10.02 | 48.0± 5.4 | SC | 10.03 | 59.9± 6.7 | SC |
| 10.41 | 36.4± 4.8 | SC | 10.41 | 36.8± 5.7 | SC |
| 10.79 | 28.7± 4.5 | SC | 10.78 | 34.1± 5.6 | SC |
| 11.16 | 22.4± 4.2 | SC | 11.14 | 28.1± 5.2 | SC |
| 11.56 | 25.0± 4.3 | SC | 11.56 | 30.0± 5.2 | SC |
| 11.92 | 29.7± 4.3 | SC | 11.92 | 32.6± 5.0 | SC |
| 12.30 | 27.4± 4.1 | SC | 12.29 | 36.1± 5.0 | SC |
| 12.68 | 23.8± 3.9 | SC | 12.67 | 27.1± 4.7 | SC |

TABLE 2 cont'd

*BV* Surface Brightness Profiles

| R (pc) | $\mu_V$ ($L_{V\odot}$ pc$^{-2}$) | Type[a] | R (pc) | $\mu_B$ ($L_{B\odot}$ pc$^{-2}$) | Type[a] |
|---|---|---|---|---|---|
| 13.04 | 21.5± 3.9 | SC | 13.04 | 28.9± 4.8 | SC |
| 13.42 | 11.3± 3.8 | SC | 13.80 | 27.4± 4.8 | SC |
| 13.81 | 16.0± 3.8 | SC | 14.19 | 25.0± 4.7 | SC |
| 14.19 | 17.4± 3.9 | SC | 14.56 | 22.8± 4.7 | SC |
| 14.56 | 11.9± 3.9 | SC | 14.94 | 7.1± 5.4 | SC |
| 14.94 | 8.9± 4.0 | SC | 15.31 | 6.8± 5.5 | SC |
| 15.31 | 4.8± 6.0 | SC | 15.71 | 12.9± 4.7 | SC |
| 15.72 | 8.1± 4.1 | SC | 16.05 | 18.4± 4.8 | SC |
| 16.07 | 12.6± 4.0 | SC | 16.46 | 15.3± 4.7 | SC |
| 16.47 | 5.1± 5.9 | SC | 16.84 | 7.8± 5.4 | SC |
| 16.83 | 3.3± 8.3 | SC | 17.22 | 17.4± 4.8 | SC |
| 17.21 | 8.9± 4.2 | SC | 17.59 | 3.4± 4.3 | SC |
| 18.27 | 4.5± 2.5 | SC | 18.18 | 6.2± 3.6 | SC |
| 19.48 | 10.3± 6.8 | SC | 18.37 | 8.6± 5.7 | SC |
| 23.53 | 3.0± 2.6 | SC | 18.71 | 7.8± 6.9 | SC |
| 27.37 | 2.1± 3.4 | SC | 19.10 | 4.9± 5.6 | SC |
| | | | 23.65 | 4.1± 4.0 | SC |
| | | | 27.37 | 3.5± 4.3 | SC |

Note: a – SP = surface photometry; SC = star counts

TABLE 3
Dependence of Radial Velocity on Position Angle

| Bin | N | Sector (deg.) | $<\Phi>$ (deg.) | Median $V_r$ (km $s^{-1}$) |
|---|---|---|---|---|
| 1 | 41 | 0– 45 | 19.7±2.0 | 495.18±0.63 |
| 2 | 40 | 45– 90 | 68.4±2.2 | 495.60±0.80 |
| 3 | 55 | 90–135 | 113.8±1.7 | 495.26±0.81 |
| 4 | 50 | 135–180 | 155.8±1.7 | 495.00±0.66 |
| 5 | 44 | 180–225 | 203.2±1.9 | 494.35±0.63 |
| 6 | 57 | 225–270 | 247.2±1.7 | 494.50±0.57 |
| 7 | 65 | 270–315 | 290.6±1.7 | 492.91±0.66 |
| 8 | 47 | 315–360 | 336.3±1.9 | 494.12±0.76 |

TABLE 4

Dependence of Apparent Rotation on Sample Size

| N | $R_{min}$ (arcmin) | A (km $s^{-1}$) | $\Phi$ (deg) |
|---|---|---|---|
| 399 | 0.0 | 1.22±0.25 | 277±12 |
| 300 | 1.5 | 1.42±0.25 | 262±10 |
| 200 | 3.1 | 1.76±0.38 | 274±13 |
| 100 | 6.2 | 1.79±0.49 | 274±17 |
| 50 | 10.0 | 2.11±0.46 | 307±13 |
| 25 | 14.3 | 1.99±1.34 | 273±38 |

TABLE 5
Binned Velocity Dispersion Profile for NGC 3201

| Bin | N | Range (pc) | $R_{med}$ (pc) | $\sigma_S$ (km $s^{-1}$) | $\sigma_{PM}$ (km $s^{-1}$) |
|---|---|---|---|---|---|
| 1 | 40 | 0.1– 1.3 | 0.960 | 3.35±0.38 | 3.35±0.40 |
| 2 | 40 | 1.3– 1.9 | 1.51 | 3.84±0.43 | 4.32±0.44 |
| 3 | 40 | 1.9– 2.6 | 2.24 | 2.99±0.34 | 2.99±0.44 |
| 4 | 40 | 2.6– 3.5 | 3.04 | 4.23±0.47 | 4.52±0.50 |
| 5 | 40 | 3.5– 4.6 | 4.13 | 3.42±0.39 | 4.27±0.44 |
| 6 | 40 | 4.6– 5.9 | 5.27 | 3.47±0.40 | 3.93±0.72 |
| 7 | 40 | 5.9– 7.7 | 6.80 | 3.09±0.39 | 3.76±0.40 |
| 8 | 40 | 7.9–10.9 | 9.29 | 3.59±0.42 | 3.61±0.45 |
| 9 | 40 | 10.9–15.9 | 12.9 | 3.02±0.35 | 3.03±0.37 |
| 10 | 39 | 16.0–47.9 | 22.5 | 1.91±0.25 | 2.00±0.26 |

TABLE 6
Adopted Mass Bins

| Bin | $M_{min}$ ($M_\odot$) | $M_{max}$ ($M_\odot$) | Contents |
|---|---|---|---|
| 1 | 0.160 | 0.250 | MS |
| 2 | 0.250 | 0.350 | MS |
| 3 | 0.350 | 0.450 | MS |
| 4 | 0.450 | 0.550 | MS & WD |
| 5 | 0.550 | 0.650 | MS |
| 6 | 0.650 | 0.750 | MS & WD |
| 7* | 0.750 | 0.826 | MS & HB & RG |
| 8 | 0.826 | 8.000 | WD |

Notes: MS = main sequence stars; WD = white dwarfs; RG = red giants; HB = horizontal branch stars
$*$ = $L_V/M$ and $L_B/M$ determined semi-empirically from CCD star counts. See text for details.

TABLE 7
$V$-Band Fitted Parameters

| $r_a/r_s$ | $x$ | $c$ | $W_0$ | $r_s$ (pc) | $\mu_0$ ($L_{V\odot}$ pc$^{-2}$) | $\chi_\nu{}^2$ ($\nu = 63$) | $P(> \chi_\nu{}^2)$ | $v_s$ (km s$^{-1}$) | Population $(M/L_V)_0$ $(M/L_V)_\odot$ | Population $(M/L_V)$ $(M/L_V)_\odot$ | Dynamical $(M/L_V)_0$ $(M/L_V)_\odot$ | Dynamical $(M/L_V)$ $(M/L_V)_\odot$ |
|---|---|---|---|---|---|---|---|---|---|---|---|---|
| $\infty$ | | 1.26 | 6.1±0.2 | 2.18±0.19 | 1797±230 | 1.075 | 0.319 | 4.50±0.17 | | | 1.62±0.11 | 1.62±0.11 |
| 10 | | 1.29 | 6.0±0.2 | 2.27±0.19 | 1730±193 | 1.077 | 0.315 | 4.52±0.18 | | | 1.62±0.11 | 1.62±0.11 |
| 5 | | 1.31 | 5.7±0.2 | 2.50±0.18 | 1553±166 | 1.094 | 0.285 | 4.62±0.18 | | | 1.67±0.12 | 1.67±0.11 |
| 3 | | 1.40 | 5.3±0.2 | 2.77±0.17 | 1427±134 | 1.111 | 0.254 | 4.78±0.19 | | | 1.69±0.14 | 1.69±0.14 |
| $\infty$ | 0.0 | 1.39 | 7.4±0.4 | 1.43±0.23 | 2136±401 | 1.110 | 0.256 | 4.16±0.17 | 4.37 | 2.88 | 2.21±0.10 | 1.45±0.14 |
| 10 | 0.0 | 1.41 | 7.2±0.2 | 1.51±0.20 | 2077±333 | 1.084 | 0.302 | 4.22±0.17 | 4.30 | 2.88 | 2.17±0.10 | 1.45±0.14 |
| 5 | 0.0 | 1.40 | 6.6±0.2 | 1.82±0.18 | 1788±227 | 1.060 | 0.348 | 4.42±0.17 | 4.03 | 2.88 | 2.18±0.10 | 1.56±0.16 |
| 3 | 0.0 | 1.43 | 5.9±0.2 | 2.15±0.16 | 1611±165 | 1.074 | 0.321 | 4.74±0.19 | 3.79 | 2.88 | 2.20±0.10 | 1.67±0.17 |
| $\infty$ | 0.5 | 1.34 | 7.6±0.4 | 1.78±0.24 | 1880±423 | 1.082 | 0.306 | 4.12±0.17 | 2.22 | 2.08 | 1.91±0.08 | 1.80±0.17 |
| 10 | 0.5 | 1.37 | 7.5±0.4 | 1.79±0.21 | 1853±271 | 1.063 | 0.343 | 4.19±0.17 | 2.19 | 2.08 | 1.91±0.08 | 1.83±0.18 |
| 5 | 0.5 | 1.42 | 7.1±0.3 | 2.00±0.18 | 1702±199 | 1.059 | 0.350 | 4.38±0.17 | 2.12 | 2.08 | 1.96±0.08 | 1.92±0.20 |
| 3 | 0.5 | 1.67 | 6.7±0.2 | 2.18±0.15 | 1678±169 | 1.095 | 0.282 | 4.64±0.18 | 2.03 | 2.08 | 1.95±0.10 | 2.00±0.24 |
| $\infty$ | 1.0 | 1.34 | 8.1±0.4 | 2.04±0.21 | 1792±237 | 1.056 | 0.357 | 4.17±0.16 | 1.29 | 1.93 | 1.58±0.07 | 2.36±0.24 |
| 10 | 1.0 | 1.40 | 8.1±0.4 | 2.09±0.20 | 1748±213 | 1.058 | 0.352 | 4.20±0.17 | 1.27 | 1.93 | 1.60±0.08 | 2.43±0.28 |
| 5 | 1.0 | 1.58 | 8.0±0.4 | 2.21±0.19 | 1657±184 | 1.086 | 0.299 | 4.32±0.17 | 1.25 | 1.93 | 1.66±0.08 | 2.56±0.36 |
| 3 | 1.0 | 1.90 | 7.3±0.2 | 2.48±0.12 | 1512± 95 | 1.129 | 0.225 | 4.60±0.18 | 1.26 | 1.93 | 1.76±0.10 | 2.69±0.34 |
| $\infty$ | 1.5 | 1.42 | 8.8±0.5 | 2.45±0.18 | 1673±170 | 1.134 | 0.218 | 4.30±0.17 | 0.84 | 2.16 | 1.36±0.07 | 3.46±0.45 |
| 10 | 1.5 | 1.55 | 8.9±0.6 | 2.46±0.21 | 1625±173 | 1.127 | 0.229 | 4.29±0.17 | 0.83 | 2.16 | 1.40±0.08 | 3.67±0.61 |
| 5 | 1.5 | 1.92 | 8.8±0.2 | 2.58±0.11 | 1503±121 | 1.128 | 0.227 | 4.36±0.17 | 0.83 | 2.16 | 1.50±0.09 | 3.87±0.49 |
| 3 | 1.5 | 2.06 | 7.4±0.1 | 3.30±0.08 | 1244± 89 | 1.218 | 0.115 | 4.85±0.19 | 0.85 | 2.16 | 1.71±0.10 | 4.36±0.55 |
| $\infty$ | 2.0 | 1.49 | 9.1±0.5 | 3.16±0.19 | 1327±108 | 1.240 | 0.095 | 4.67±0.18 | 0.66 | 2.69 | 1.42±0.08 | 5.84±1.06 |
| 10 | 2.0 | 1.51 | 8.6±0.5 | 3.41±0.18 | 1201± 88 | 1.222 | 0.111 | 4.82±0.19 | 0.68 | 2.69 | 1.53±0.09 | 6.07±1.00 |
| 5 | 2.0 | 1.73 | 8.2±0.3 | 3.69±0.15 | 1120± 68 | 1.219 | 0.114 | 5.07±0.20 | 0.69 | 2.69 | 1.64±0.10 | 6.26±0.80 |
| 3 | 2.0 | 1.78 | 7.1±0.1 | 4.07±0.15 | 1050± 57 | 1.208 | 0.124 | 5.54±0.22 | 0.77 | 2.69 | 1.80±0.10 | 6.38±0.62 |

TABLE 8

$B$-Band Fitted Parameters

| $r_a/r_s$ | $x$ | $c$ | $W_0$ | $r_s$ (pc) | $\mu_0$ ($L_{B\odot}$ pc$^{-2}$) | ${\chi_\nu}^2$ ($\nu = 65$) | $P(> {\chi_\nu}^2)$ | $v_s$ (km s$^{-1}$) | Population $(M/L_B)_0$ $(M/L_B)_\odot$ | Population $(M/L_B)$ $(M/L_B)_\odot$ | Dynamical $(M/L_B)_0$ $(M/L_B)_\odot$ | Dynamical $(M/L_B)$ $(M/L_B)_\odot$ |
|---|---|---|---|---|---|---|---|---|---|---|---|---|
| $\infty$ | | 1.33 | 6.3±0.2 | 2.08±0.17 | 1822±206 | 0.955 | 0.581 | 4.46±0.17 | | | 1.66±0.11 | 1.66±0.11 |
| 10 | | 1.35 | 6.2±0.2 | 2.15±0.16 | 1768±199 | 0.960 | 0.568 | 4.50±0.18 | | | 1.66±0.11 | 1.66±0.11 |
| 5 | | 1.39 | 5.9±0.2 | 2.38±0.16 | 1596±177 | 0.994 | 0.490 | 4.59±0.18 | | | 1.70±0.13 | 1.70±0.13 |
| 3 | | 1.56 | 5.6±0.2 | 2.64±0.10 | 1468±112 | 1.041 | 0.386 | 4.73±0.19 | | | 1.70±0.13 | 1.70±0.13 |
| $\infty$ | 0.0 | 1.45 | 7.7±0.3 | 1.31±0.19 | 2184±344 | 1.010 | 0.454 | 4.10±0.17 | 4.89 | 3.09 | 2.32±0.11 | 1.47±0.14 |
| 10 | 0.0 | 1.48 | 7.5±0.3 | 1.39±0.16 | 2134±292 | 0.970 | 0.545 | 4.19±0.16 | 4.80 | 3.09 | 2.30±0.11 | 1.48±0.14 |
| 5 | 0.0 | 1.51 | 6.9±0.2 | 1.65±0.15 | 1890±217 | 0.957 | 0.575 | 4.38±0.18 | 4.51 | 3.09 | 2.27±0.11 | 1.55±0.16 |
| 3 | 0.0 | 1.61 | 6.2±0.2 | 1.97±0.13 | 1710±167 | 0.972 | 0.541 | 4.67±0.19 | 4.24 | 3.09 | 2.24±0.11 | 1.64±0.13 |
| $\infty$ | 0.5 | 1.42 | 8.1±0.4 | 1.50±0.19 | 2057±294 | 0.977 | 0.529 | 3.99±0.16 | 2.47 | 2.25 | 1.93±0.10 | 1.76±0.17 |
| 10 | 0.5 | 1.46 | 7.9±0.3 | 1.60±0.17 | 1969±252 | 0.957 | 0.575 | 4.08±0.16 | 2.45 | 2.25 | 1.95±0.10 | 1.81±0.17 |
| 5 | 0.5 | 1.58 | 7.5±0.3 | 1.78±0.16 | 1835±208 | 0.961 | 0.566 | 4.26±0.17 | 2.35 | 2.25 | 1.99±0.10 | 1.90±0.21 |
| 3 | 0.5 | 1.72 | 6.8±0.2 | 2.16±0.14 | 1616±153 | 1.004 | 0.468 | 4.60±0.18 | 2.17 | 2.25 | 2.01±0.10 | 2.07±0.25 |
| $\infty$ | 1.0 | 1.41 | 8.6±0.4 | 1.88±0.18 | 1855±216 | 0.944 | 0.605 | 4.02±0.15 | 1.38 | 2.08 | 1.58±0.08 | 2.38±0.25 |
| 10 | 1.0 | 1.47 | 8.5±0.4 | 1.96±0.18 | 1769±195 | 0.958 | 0.573 | 4.08±0.16 | 1.36 | 2.08 | 1.62±0.09 | 2.47±0.30 |
| 5 | 1.0 | 1.73 | 8.3±0.4 | 2.13±0.17 | 1640±166 | 0.996 | 0.486 | 4.22±0.17 | 1.29 | 2.08 | 1.68±0.11 | 2.71±0.38 |
| 3 | 1.0 | 2.14 | 7.5±0.1 | 2.47±0.13 | 1570±164 | 1.046 | 0.376 | 4.53±0.18 | 1.27 | 2.08 | 1.78±0.13 | 2.84±0.41 |
| $\infty$ | 1.5 | 1.41 | 9.3±0.4 | 2.34±0.17 | 1648±152 | 1.018 | 0.436 | 4.17±0.16 | 0.88 | 2.35 | 1.38±0.08 | 3.65±0.49 |
| 10 | 1.5 | 1.61 | 9.3±0.4 | 2.41±0.18 | 1585±148 | 1.024 | 0.423 | 4.20±0.16 | 0.87 | 2.35 | 1.43±0.08 | 3.85±0.62 |
| 5 | 1.5 | 1.95 | 8.8±0.2 | 2.61±0.11 | 1440±106 | 1.067 | 0.333 | 4.35±0.17 | 0.89 | 2.35 | 1.55±0.10 | 4.10±0.52 |
| 3 | 1.5 | 2.12 | 7.4±0.1 | 3.39±0.12 | 1167± 85 | 1.155 | 0.184 | 4.83±0.19 | 0.80 | 2.35 | 1.80±0.10 | 4.70±0.55 |
| $\infty$ | 2.0 | 1.59 | 9.8±0.4 | 2.91±0.16 | 1390±116 | 1.145 | 0.198 | 4.47±0.17 | 0.67 | 2.95 | 1.40±0.07 | 6.14±1.10 |
| 10 | 2.0 | 1.60 | 9.1±0.3 | 3.26±0.15 | 1205± 88 | 1.173 | 0.161 | 4.69±0.18 | 0.70 | 2.95 | 1.55±0.08 | 6.45±0.95 |
| 5 | 2.0 | 1.76 | 8.3±0.3 | 3.61±0.15 | 1101± 66 | 1.230 | 0.100 | 5.03±0.20 | 0.74 | 2.95 | 1.68±0.10 | 6.81±0.82 |
| 3 | 2.0 | 1.60 | 6.8±0.2 | 4.16±0.12 | 979± 51 | 1.371 | 0.025 | 5.61±0.22 | 0.86 | 2.95 | 1.93±0.10 | 6.54±0.74 |

TABLE 9

*V*-Band Derived Parameters

| $r_a/r_s$ | $x$ | $r_s$ (pc) | $r_h$ (pc) | $r_t$ (pc) | $v_s\eta_0$ (km s$^{-1}$) | L$_V$ ($10^4$L$_{V\odot}$) | $\log\sigma_0$ (L$_{V\odot}$ pc$^{-3}$) | M ($10^5$M$_\odot$) | $\log\rho_0$ (M$_\odot$ pc$^{-3}$) | $\log\rho_h$ (M$_\odot$ pc$^{-3}$) | $\log\rho_t$ (M$_\odot$ pc$^{-3}$) | $\log t_{r0}$ (yr) | $\log t_{rh}$ (yr) |
|---|---|---|---|---|---|---|---|---|---|---|---|---|---|
| $\infty$ | | 2.18±0.19 | 6.2±0.6 | 39.9± 2.3 | 4.22±0.17 | 8.00±0.22 | 2.64±0.09 | 1.30±0.08 | 2.85±0.08 | 1.83±0.08 | -0.31±0.07 | 8.01±0.16 | 8.79±0.05 |
| 10 | | 2.27±0.19 | 6.2±0.6 | 43.7± 3.4 | 4.25±0.18 | 8.02±0.23 | 2.61±0.08 | 1.30±0.08 | 2.82±0.07 | 1.82±0.09 | -0.43±0.10 | 8.04±0.16 | 8.80±0.05 |
| 5 | | 2.50±0.18 | 6.3±0.8 | 50.8± 7.6 | 4.30±0.18 | 8.01±0.27 | 2.53±0.07 | 1.34±0.09 | 2.75±0.07 | 1.81±0.10 | -0.61±0.18 | 8.14±0.12 | 8.84±0.04 |
| 3 | | 2.77±0.17 | 6.5±0.8 | 70.1± 31.1 | 4.40±0.19 | 8.15±0.36 | 2.47±0.06 | 1.38±0.10 | 2.70±0.06 | 1.78±0.11 | -1.02±0.41 | 8.24±0.09 | 8.89±0.03 |
| $\infty$ | 0.0 | 1.43±0.23 | 6.9±0.5 | 35.0± 1.0 | 5.04±0.17 | 7.87±0.22 | 2.81±0.14 | 1.14±0.08 | 3.15±0.16 | 1.63±0.10 | -0.20±0.04 | 7.53±0.11 | 8.83±0.04 |
| 10 | 0.0 | 1.51±0.20 | 5.7±0.5 | 38.7± 2.0 | 5.09±0.17 | 7.98±0.24 | 2.78±0.12 | 1.16±0.08 | 3.11±0.11 | 1.87±0.10 | -0.32±0.07 | 7.58±0.09 | 8.75±0.05 |
| 5 | 0.0 | 1.82±0.18 | 6.9±0.6 | 45.8± 5.0 | 5.19±0.17 | 8.01±0.27 | 2.66±0.09 | 1.25±0.08 | 2.99±0.08 | 1.66±0.10 | -0.51±0.14 | 7.76±0.07 | 8.84±0.05 |
| 3 | 0.0 | 2.15±0.16 | 8.1±0.8 | 57.6± 11.1 | 5.37±0.19 | 8.12±0.30 | 2.57±0.07 | 1.35±0.09 | 2.91±0.06 | 1.48±0.14 | -0.77±0.22 | 7.93±0.06 | 8.93±0.05 |
| $\infty$ | 0.5 | 1.73±0.24 | 6.5±0.6 | 37.4± 1.3 | 4.66±0.17 | 7.84±0.22 | 2.70±0.12 | 1.41±0.09 | 2.98±0.12 | 1.78±0.10 | -0.20±0.03 | 7.76±0.09 | 8.94±0.05 |
| 10 | 0.5 | 1.79±0.21 | 6.8±0.6 | 41.6± 2.6 | 4.73±0.17 | 7.95±0.24 | 2.68±0.11 | 1.45±0.09 | 2.96±0.10 | 1.75±0.09 | -0.32±0.07 | 7.80±0.08 | 8.96±0.05 |
| 5 | 0.5 | 2.00±0.18 | 7.6±0.6 | 52.5± 7.9 | 4.87±0.17 | 8.08±0.28 | 2.61±0.08 | 1.55±0.10 | 2.90±0.08 | 1.63±0.09 | -0.59±0.17 | 7.91±0.07 | 9.02±0.04 |
| 3 | 0.5 | 2.18±0.15 | 8.0±0.8 | 98.5± 43.8 | 5.11±0.19 | 8.61±0.40 | 2.59±0.07 | 1.72±0.12 | 2.88±0.06 | 1.60±0.11 | -1.37±0.41 | 8.01±0.06 | 9.06±0.06 |
| $\infty$ | 1.0 | 2.04±0.21 | 9.8±0.7 | 45.0± 2.1 | 4.28±0.16 | 8.04±0.23 | 2.64±0.09 | 1.90±0.13 | 2.84±0.09 | 1.39±0.09 | -0.30±0.04 | 7.98±0.07 | 9.24±0.04 |
| 10 | 1.0 | 2.09±0.20 | 10.0±0.7 | 52.2± 4.8 | 4.33±0.17 | 8.14±0.25 | 2.62±0.09 | 1.98±0.15 | 2.83±0.09 | 1.37±0.09 | -0.48±0.10 | 8.01±0.07 | 9.26±0.04 |
| 5 | 1.0 | 2.21±0.19 | 10.6±1.0 | 84.1± 36.6 | 4.50±0.17 | 8.41±0.37 | 2.58±0.08 | 2.16±0.20 | 2.80±0.08 | 1.34±0.10 | -1.06±0.37 | 8.06±0.07 | 9.30±0.06 |
| 3 | 1.0 | 2.48±0.12 | 11.9±1.1 | 195.2± 98.6 | 4.75±0.18 | 8.84±0.29 | 2.51±0.05 | 2.37±0.20 | 2.76±0.04 | 1.23±0.09 | -2.12±0.47 | 8.19±0.05 | 9.37±0.06 |
| $\infty$ | 1.5 | 2.45±0.18 | 11.7±0.9 | 64.8± 4.8 | 3.95±0.17 | 8.39±0.24 | 2.58±0.07 | 2.91±0.26 | 2.71±0.06 | 1.33±0.08 | -0.59±0.07 | 8.20±0.06 | 9.45±0.05 |
| 10 | 1.5 | 2.46±0.21 | 14.2±1.4 | 86.5± 18.7 | 4.01±0.17 | 8.51±0.31 | 2.56±0.07 | 3.13±0.35 | 2.71±0.07 | 1.11±0.09 | -0.94±0.21 | 8.21±0.07 | 9.55±0.06 |
| 5 | 1.5 | 2.58±0.11 | 14.9±1.7 | 216.5± 86.7 | 4.20±0.17 | 8.74±0.22 | 2.50±0.04 | 3.38±0.29 | 2.68±0.04 | 1.08±0.11 | -2.10±0.39 | 8.25±0.05 | 9.59±0.07 |
| 3 | 1.5 | 3.30±0.08 | 15.8±2.1 | 377.2±148.2 | 4.34±0.19 | 8.51±0.16 | 2.32±0.05 | 3.71±0.32 | 2.56±0.03 | 1.05±0.13 | -2.80±0.44 | 8.51±0.05 | 9.63±0.08 |
| $\infty$ | 2.0 | 3.16±0.19 | 18.2±1.9 | 97.9± 12.0 | 3.79±0.18 | 8.40±0.28 | 2.41±0.05 | 4.90±0.61 | 2.56±0.05 | 0.98±0.09 | -0.90±0.11 | 8.49±0.07 | 9.78±0.07 |
| 10 | 2.0 | 3.41±0.18 | 16.3±2.0 | 110.6± 22.6 | 3.85±0.19 | 8.15±0.22 | 2.34±0.05 | 4.95±0.59 | 2.52±0.05 | 1.13±0.11 | -1.06±0.20 | 8.57±0.06 | 9.73±0.08 |
| 5 | 2.0 | 3.69±0.15 | 21.3±1.6 | 199.7± 47.5 | 3.97±0.20 | 8.13±0.16 | 2.29±0.04 | 5.38±0.44 | 2.50±0.04 | 0.82±0.09 | -1.79±0.26 | 8.67±0.05 | 9.87±0.05 |
| 3 | 2.0 | 4.07±0.15 | 19.4±1.7 | 244.3± 42.0 | 4.17±0.22 | 8.02±0.14 | 2.24±0.04 | 5.12±0.34 | 2.49±0.03 | 0.92±0.12 | -2.08±0.20 | 8.79±0.04 | 9.82±0.05 |

TABLE 10
*B*-Band Derived Parameters

| $r_a/r_s$ | $x$ | $r_s$ (pc) | $r_h$ (pc) | $r_t$ (pc) | $v_s\eta_0$ (km s$^{-1}$) | L$_B$ ($10^4$L$_{B\odot}$) | $\log\sigma_0$ (L$_{B\odot}$ pc$^{-3}$) | M ($10^5$M$_\odot$) | $\log\rho_0$ (M$_\odot$ pc$^{-3}$) | $\log\rho_h$ (M$_\odot$ pc$^{-3}$) | $\log\rho_t$ (M$_\odot$ pc$^{-3}$) | $\log t_{r0}$ (yr) | $\log t_{rh}$ (yr) |
|---|---|---|---|---|---|---|---|---|---|---|---|---|---|
| ∞ | | 2.08±0.17 | 6.3±0.5 | 44.2± 2.6 | 4.22±0.17 | 7.86±0.23 | 2.67±0.08 | 1.31±0.08 | 2.89±0.07 | 1.81±0.07 | -0.44±0.08 | 7.96±0.16 | 8.76±0.05 |
| 10 | | 2.15±0.16 | 6.3±0.5 | 47.9± 4.3 | 4.25±0.18 | 7.91±0.25 | 2.64±0.08 | 1.32±0.08 | 2.86±0.07 | 1.81±0.07 | -0.55±0.11 | 8.00±0.15 | 8.78±0.05 |
| 5 | | 2.38±0.16 | 6.5±0.5 | 57.8± 16.2 | 4.31±0.18 | 7.94±0.35 | 2.56±0.07 | 1.35±0.09 | 2.79±0.06 | 1.78±0.07 | -0.78±0.28 | 8.09±0.11 | 8.82±0.04 |
| 3 | | 2.64±0.10 | 6.7±0.6 | 95.8± 47.2 | 4.41±0.19 | 8.19±0.30 | 2.50±0.05 | 1.39±0.09 | 2.73±0.04 | 1.74±0.07 | -1.42±0.40 | 8.19±0.06 | 8.87±0.02 |
| ∞ | 0.0 | 1.31±0.19 | 6.3±0.4 | 36.6± 1.1 | 5.05±0.17 | 7.67±0.19 | 2.85±0.12 | 1.13±0.07 | 3.21±0.12 | 1.74±0.08 | -0.26±0.10 | 7.45±0.10 | 8.79±0.04 |
| 10 | 0.0 | 1.39±0.16 | 6.7±0.5 | 41.6± 2.5 | 5.13±0.16 | 7.83±0.22 | 2.82±0.10 | 1.16±0.07 | 3.18±0.10 | 1.67±0.09 | -0.42±0.08 | 7.51±0.08 | 8.82±0.04 |
| 5 | 0.0 | 1.65±0.15 | 6.3±0.5 | 53.7± 7.8 | 5.24±0.18 | 7.99±0.28 | 2.71±0.08 | 1.24±0.08 | 3.07±0.08 | 1.78±0.10 | -0.72±0.17 | 7.67±0.06 | 8.80±0.05 |
| 3 | 0.0 | 1.97±0.13 | 7.3±0.5 | 79.3± 25.4 | 5.43±0.19 | 8.24±0.36 | 2.62±0.07 | 1.36±0.09 | 2.97±0.06 | 1.63±0.09 | -1.19±0.33 | 7.85±0.05 | 8.88±0.04 |
| ∞ | 0.5 | 1.50±0.19 | 7.2±0.5 | 39.6± 1.3 | 4.66±0.16 | 7.75±0.19 | 2.79±0.11 | 1.36±0.09 | 3.07±0.11 | 1.64±0.08 | -0.28±0.03 | 7.62±0.09 | 8.97±0.04 |
| 10 | 0.5 | 1.60±0.17 | 7.6±0.5 | 45.6± 3.3 | 4.74±0.16 | 7.86±0.23 | 2.75±0.09 | 1.42±0.09 | 3.04±0.09 | 1.58±0.08 | -0.45±0.08 | 7.69±0.07 | 9.00±0.04 |
| 5 | 0.5 | 1.78±0.16 | 8.5±0.6 | 68.2± 17.4 | 4.91±0.17 | 8.16±0.33 | 2.68±0.08 | 1.55±0.11 | 2.98±0.08 | 1.48±0.09 | -0.93±0.27 | 7.80±0.07 | 9.07±0.05 |
| 3 | 0.5 | 2.16±0.14 | 10.1±0.9 | 113.3± 54.8 | 5.09±0.18 | 8.40±0.40 | 2.57±0.06 | 1.74±0.13 | 2.88±0.06 | 1.30±0.11 | -1.55±0.45 | 7.99±0.06 | 9.16±0.05 |
| ∞ | 1.0 | 1.88±0.18 | 9.0±0.6 | 48.4± 2.3 | 4.25±0.15 | 7.92±0.21 | 2.89±0.08 | 1.89±0.13 | 2.89±0.08 | 1.49±0.08 | -0.40±0.04 | 7.89±0.07 | 9.20±0.04 |
| 10 | 1.0 | 1.96±0.18 | 9.4±0.7 | 58.0± 6.4 | 4.32±0.16 | 8.01±0.24 | 2.86±0.08 | 1.98±0.15 | 2.86±0.08 | 1.46±0.08 | -0.62±0.11 | 7.93±0.07 | 9.23±0.05 |
| 5 | 1.0 | 2.13±0.17 | 12.3±1.2 | 115.6± 52.6 | 4.47±0.17 | 8.35±0.36 | 2.59±0.07 | 2.26±0.21 | 2.82±0.07 | 1.16±0.10 | -1.46±0.43 | 8.02±0.07 | 9.37±0.06 |
| 3 | 1.0 | 2.47±0.13 | 14.3±1.3 | 341.6±140.0 | 4.71±0.18 | 8.80±0.18 | 2.50±0.06 | 2.50±0.06 | 2.75±0.04 | 1.01±0.09 | -2.82±0.40 | 8.17±0.06 | 9.47±0.06 |
| ∞ | 1.5 | 2.34±0.17 | 13.6±1.0 | 71.9± 5.5 | 3.92±0.16 | 8.25±0.24 | 2.58±0.06 | 3.01±0.27 | 2.72±0.06 | 1.16±0.08 | -0.71±0.07 | 8.15±0.06 | 9.52±0.05 |
| 10 | 1.5 | 2.41±0.18 | 13.9±1.5 | 97.3± 22.6 | 4.00±0.16 | 8.28±0.29 | 2.55±0.07 | 3.19±0.35 | 2.71±0.07 | 1.15±0.09 | -1.08±0.22 | 8.18±0.07 | 9.55±0.07 |
| 5 | 1.5 | 2.61±0.11 | 17.7±1.7 | 243.3± 85.7 | 4.19±0.17 | 8.39±0.20 | 2.48±0.04 | 3.44±0.30 | 2.67±0.04 | 0.87±0.11 | -2.20±0.37 | 8.26±0.05 | 9.66±0.06 |
| 3 | 1.5 | 3.39±0.12 | 16.2±1.1 | 447.7±140.5 | 4.32±0.19 | 8.13±0.17 | 2.28±0.04 | 3.82±0.31 | 2.53±0.03 | 1.03±0.07 | -3.00±0.41 | 8.53±0.04 | 9.65±0.04 |
| ∞ | 2.0 | 2.91±0.16 | 19.7±2.1 | 112.9± 13.8 | 3.77±0.17 | 8.35±0.29 | 2.45±0.06 | 5.13±0.63 | 2.60±0.05 | 0.90±0.09 | -1.07±0.10 | 8.40±0.07 | 9.82±0.07 |
| 10 | 2.0 | 3.26±0.15 | 18.8±1.9 | 130.7± 27.1 | 3.84±0.18 | 7.95±0.20 | 2.35±0.05 | 5.12±0.52 | 2.54±0.04 | 0.96±0.09 | -1.26±0.19 | 8.52±0.06 | 9.80±0.06 |
| 5 | 2.0 | 3.61±0.15 | 20.9±1.7 | 208.3± 51.0 | 3.98±0.20 | 7.79±0.15 | 2.29±0.04 | 5.31±0.44 | 2.51±0.04 | 0.84±0.09 | -1.85±0.26 | 8.64±0.05 | 9.85±0.05 |
| 3 | 2.0 | 4.16±0.12 | 15.7±1.3 | 165.7± 36.0 | 4.18±0.22 | 7.42±0.14 | 2.20±0.03 | 4.86±0.38 | 2.48±0.02 | 1.17±0.09 | -1.60±0.24 | 8.81±0.04 | 9.71±0.05 |

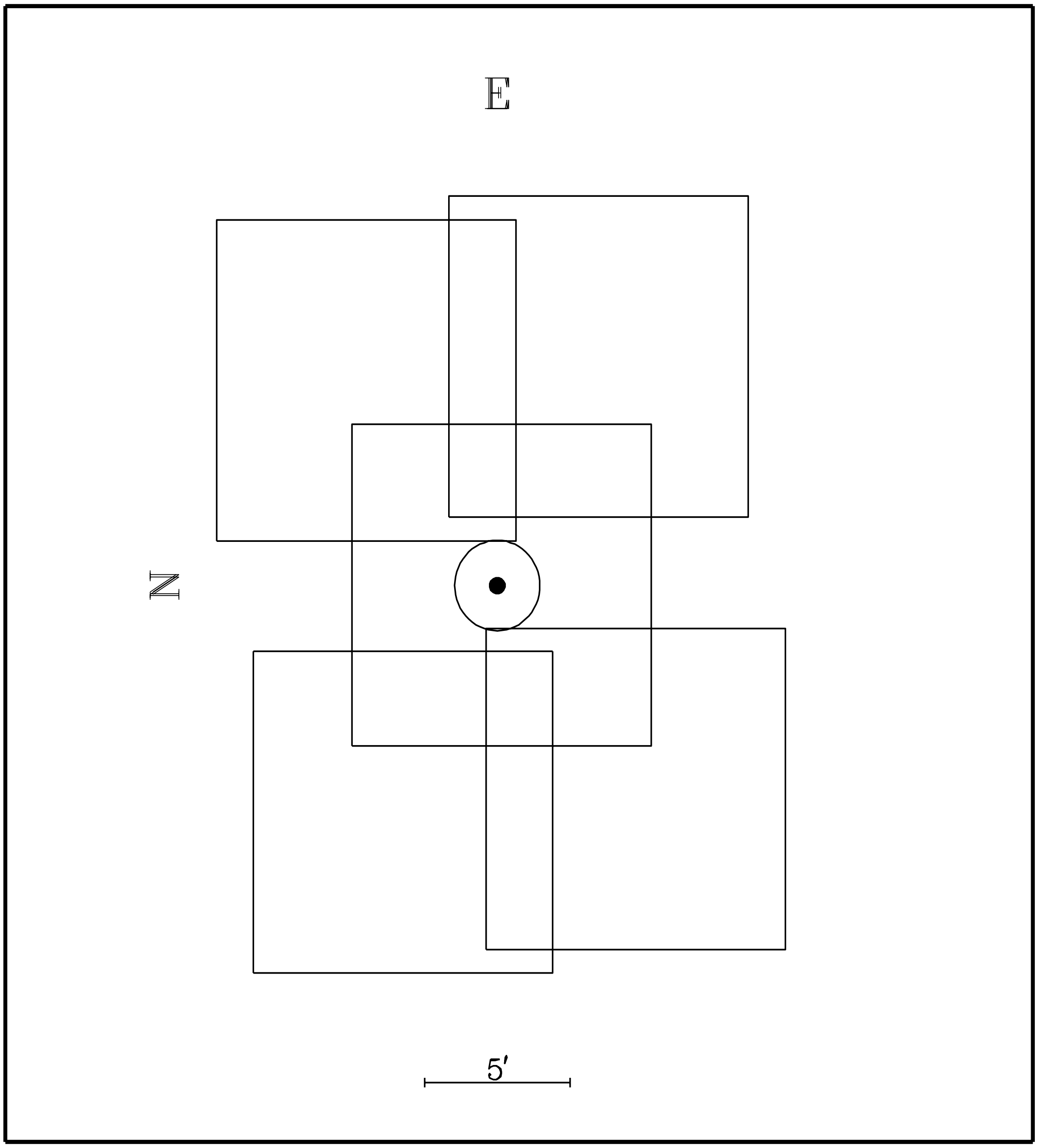

Figure 1

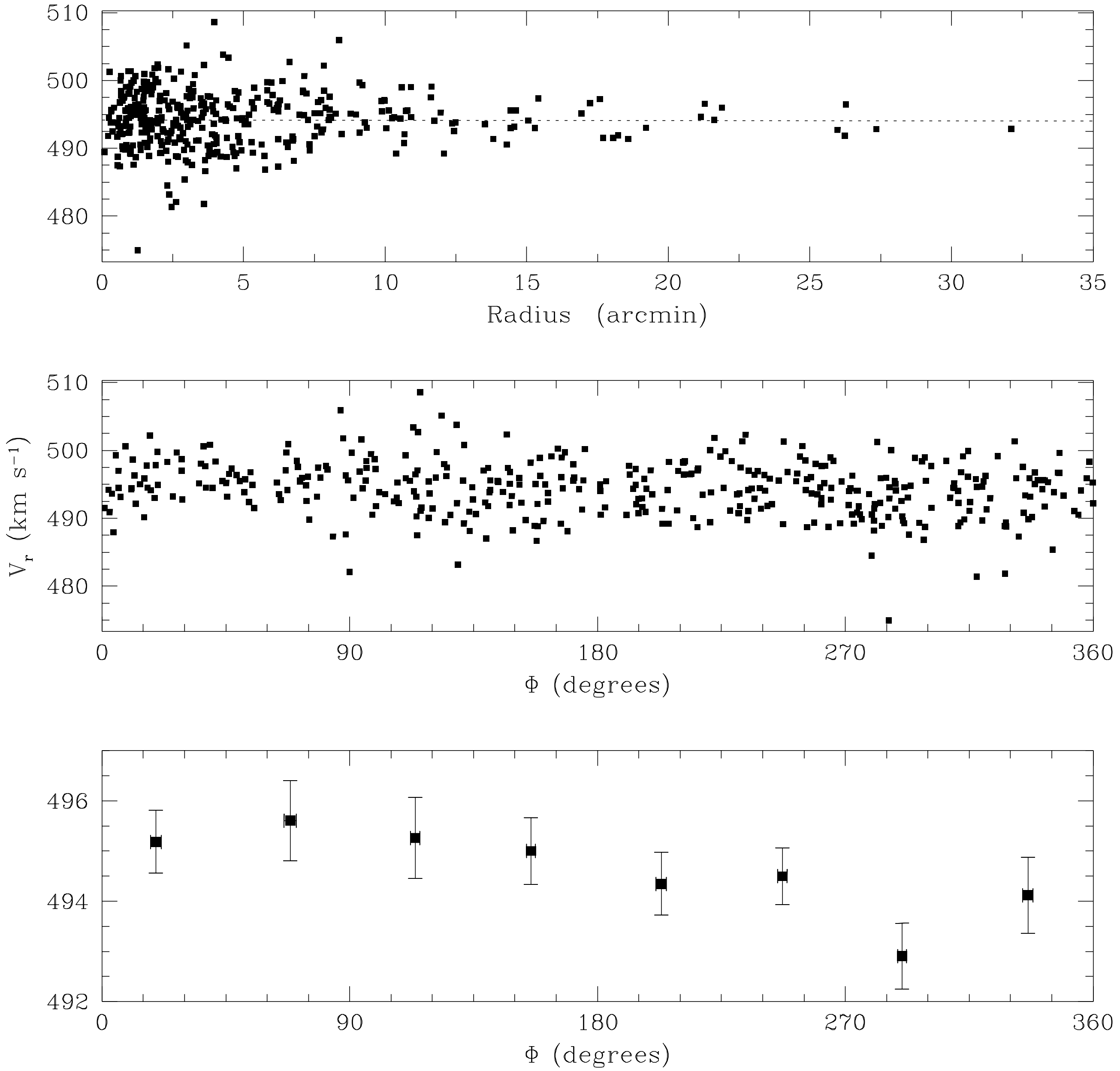

**Figure 2**

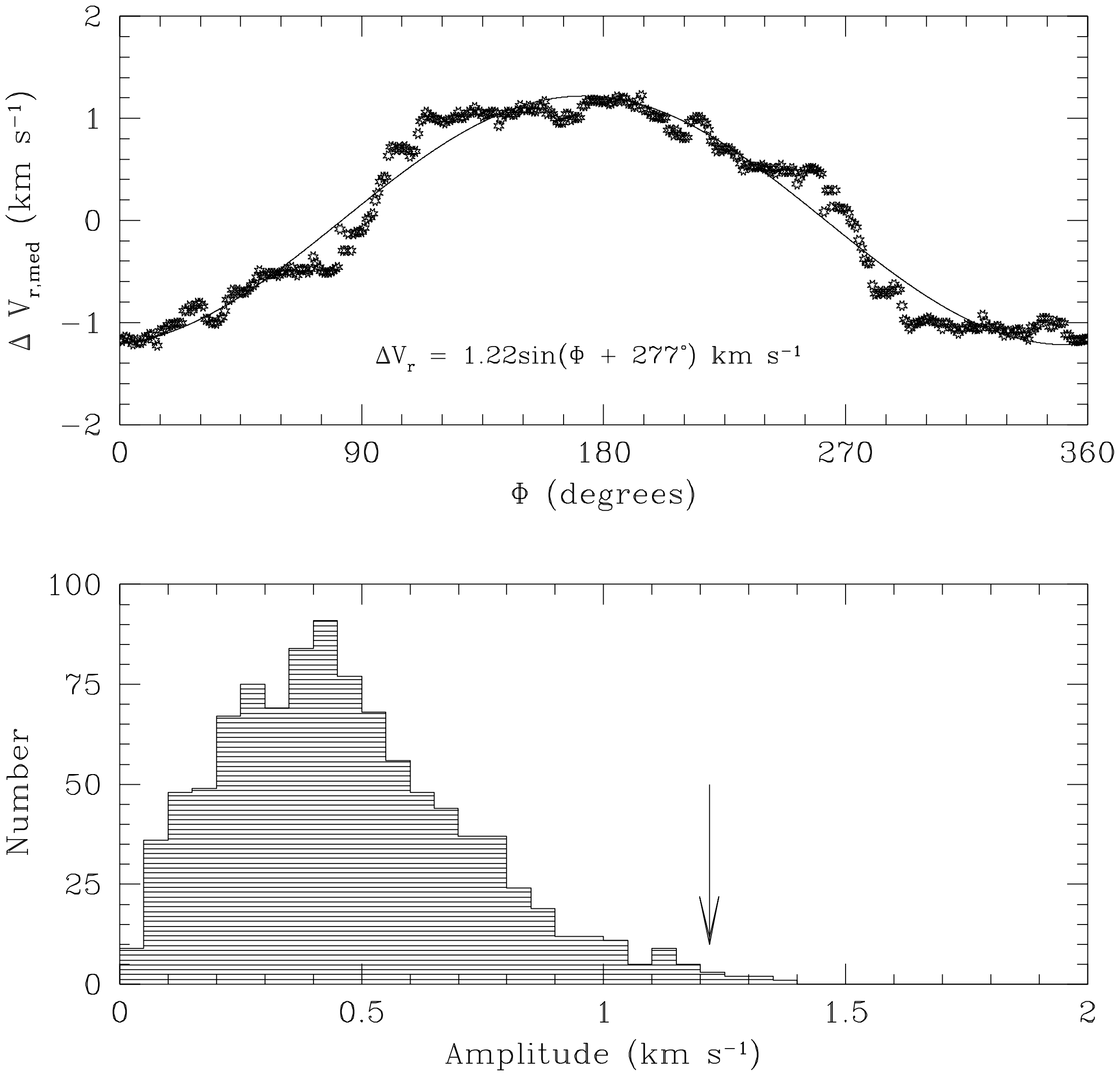

**Figure 3**

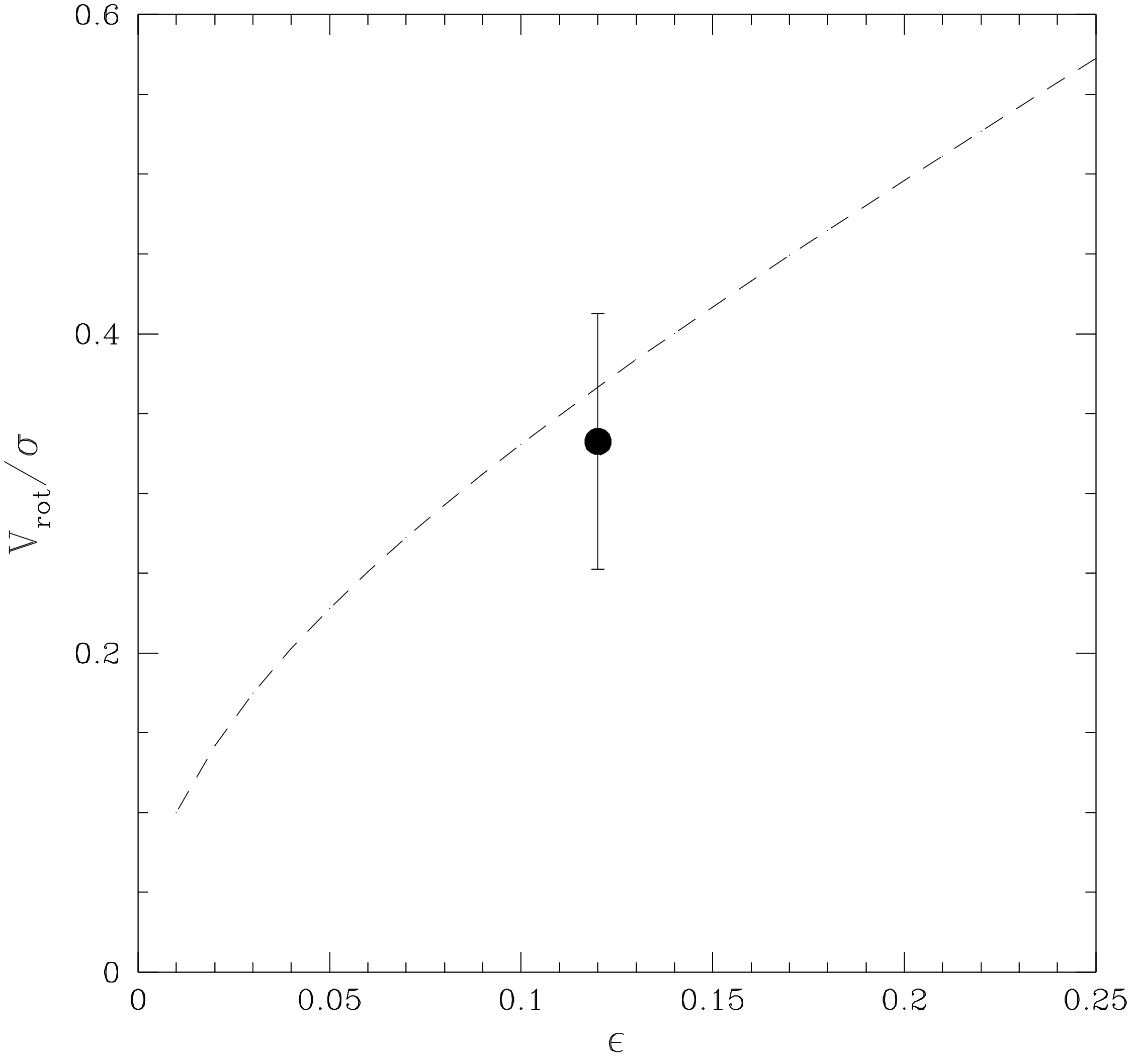

**Figure 4**

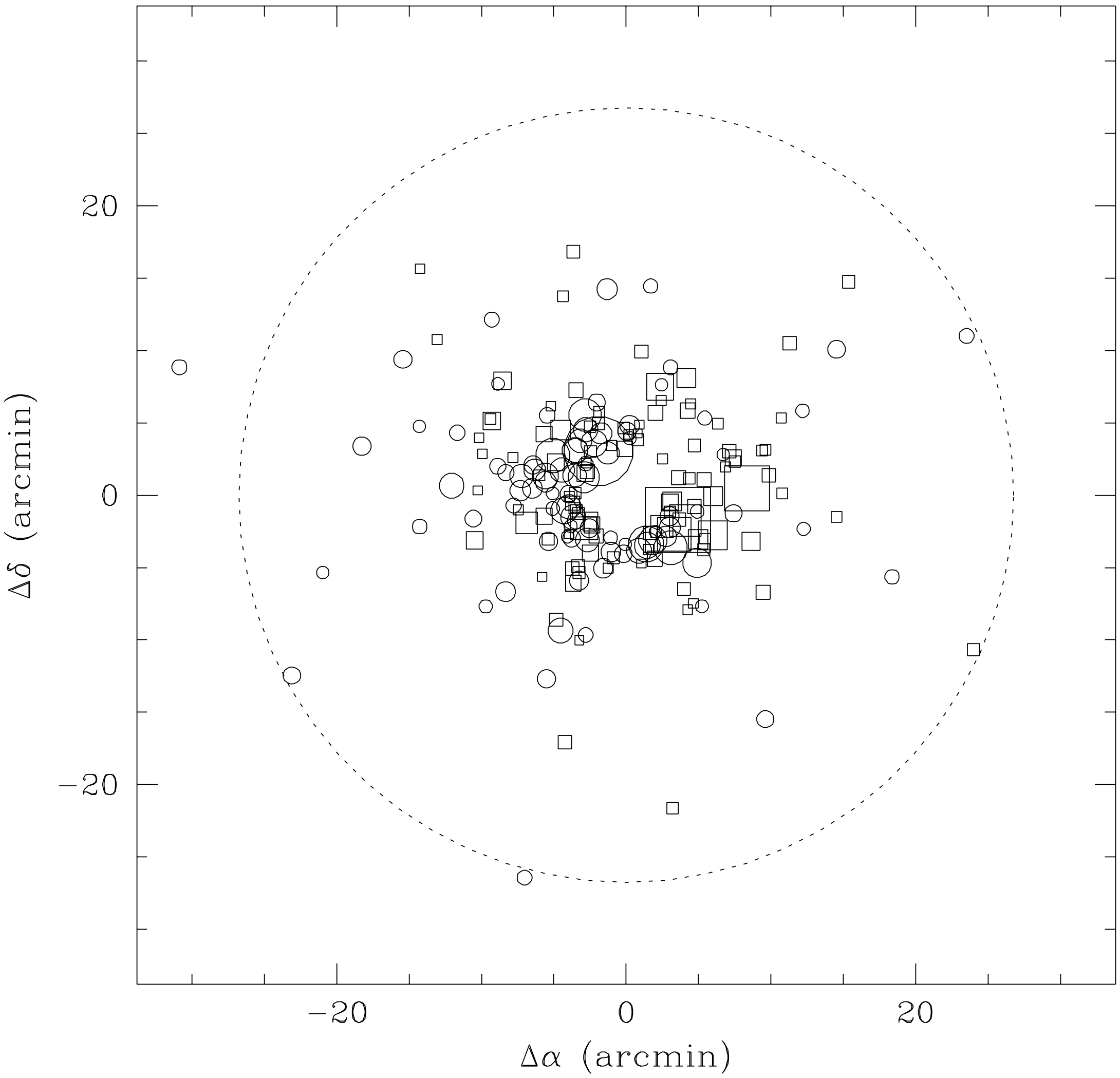

**Figure 5**

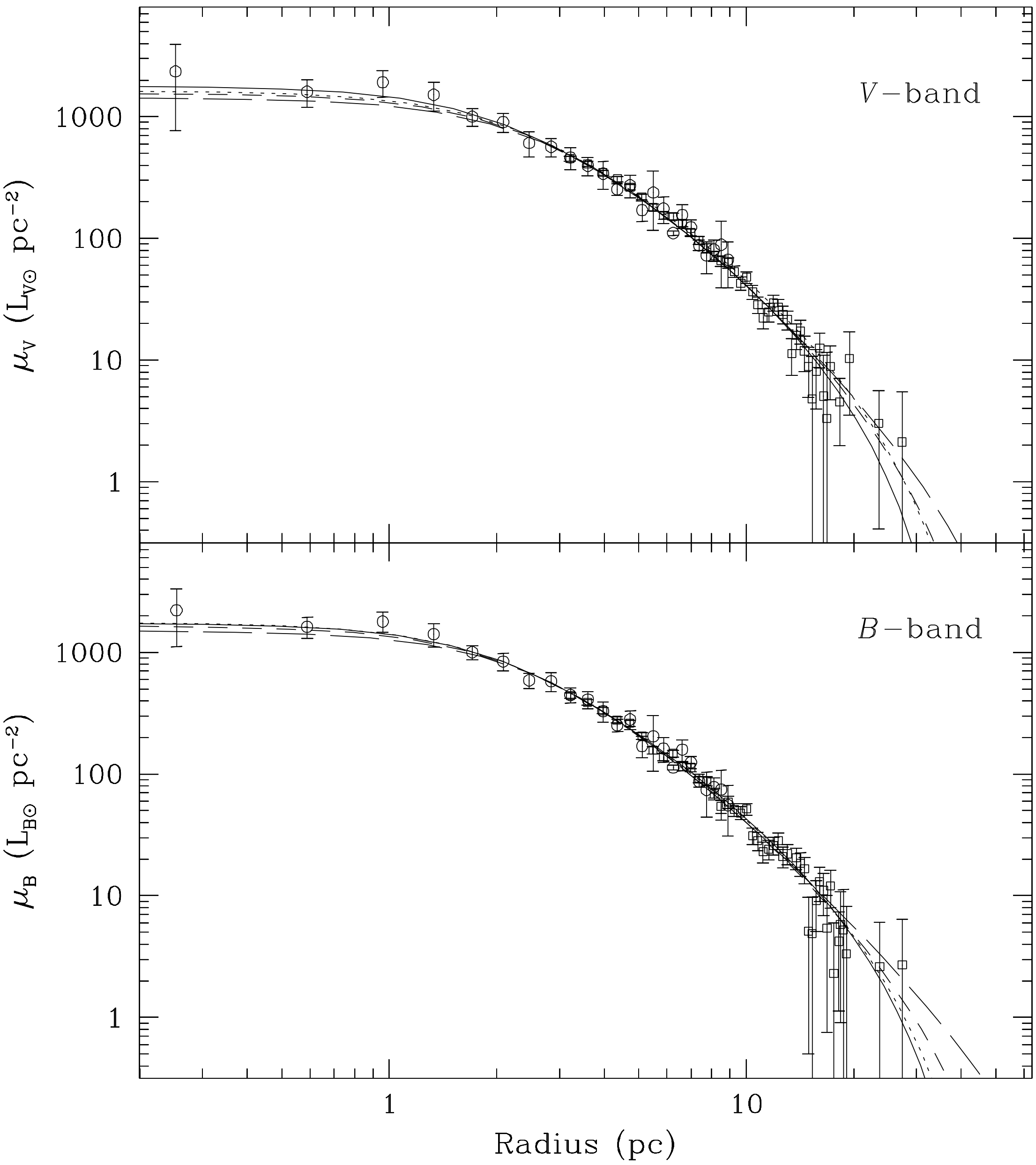

**Figure 6**

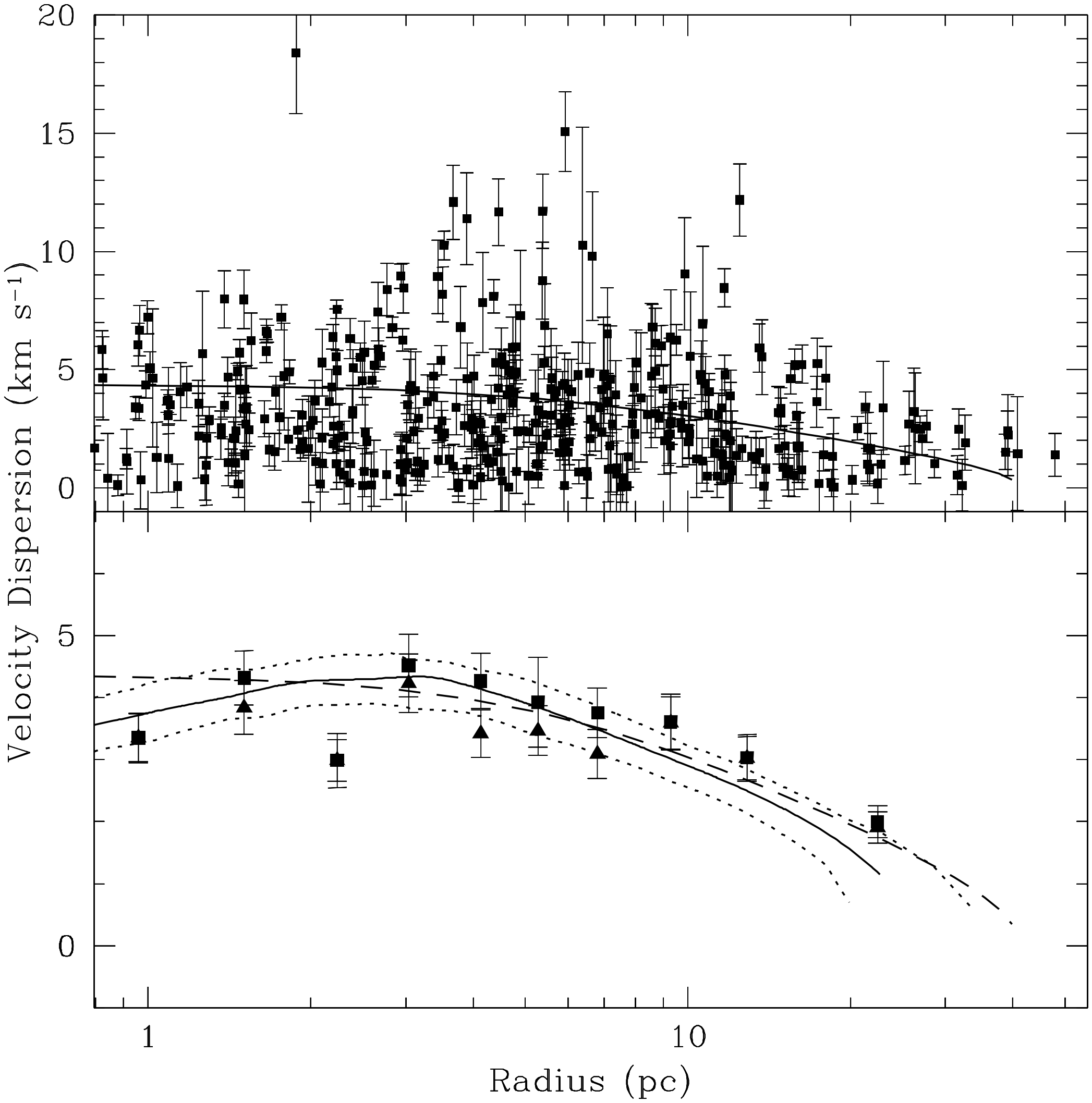

**Figure 7**

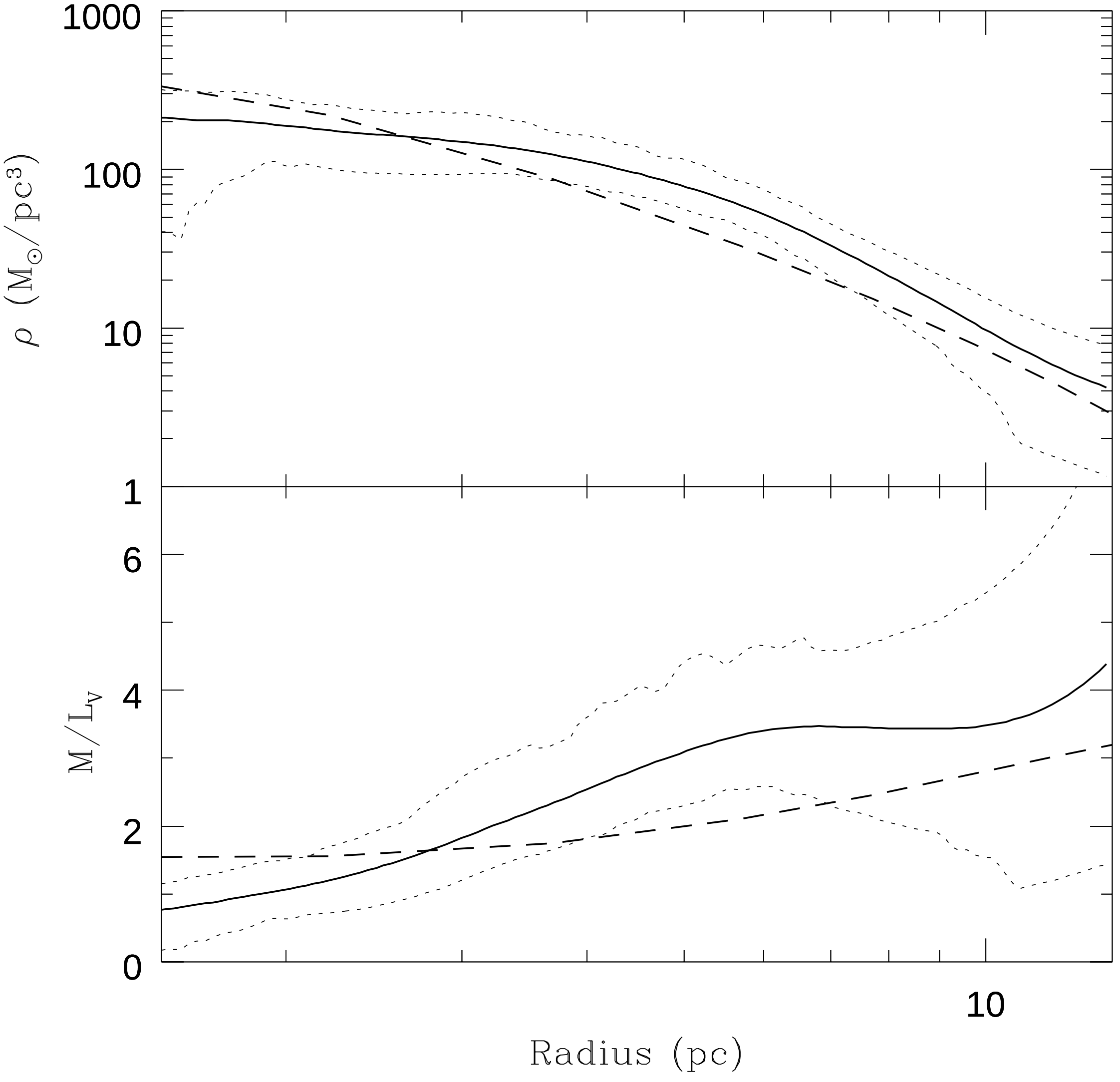

**Figure 8**

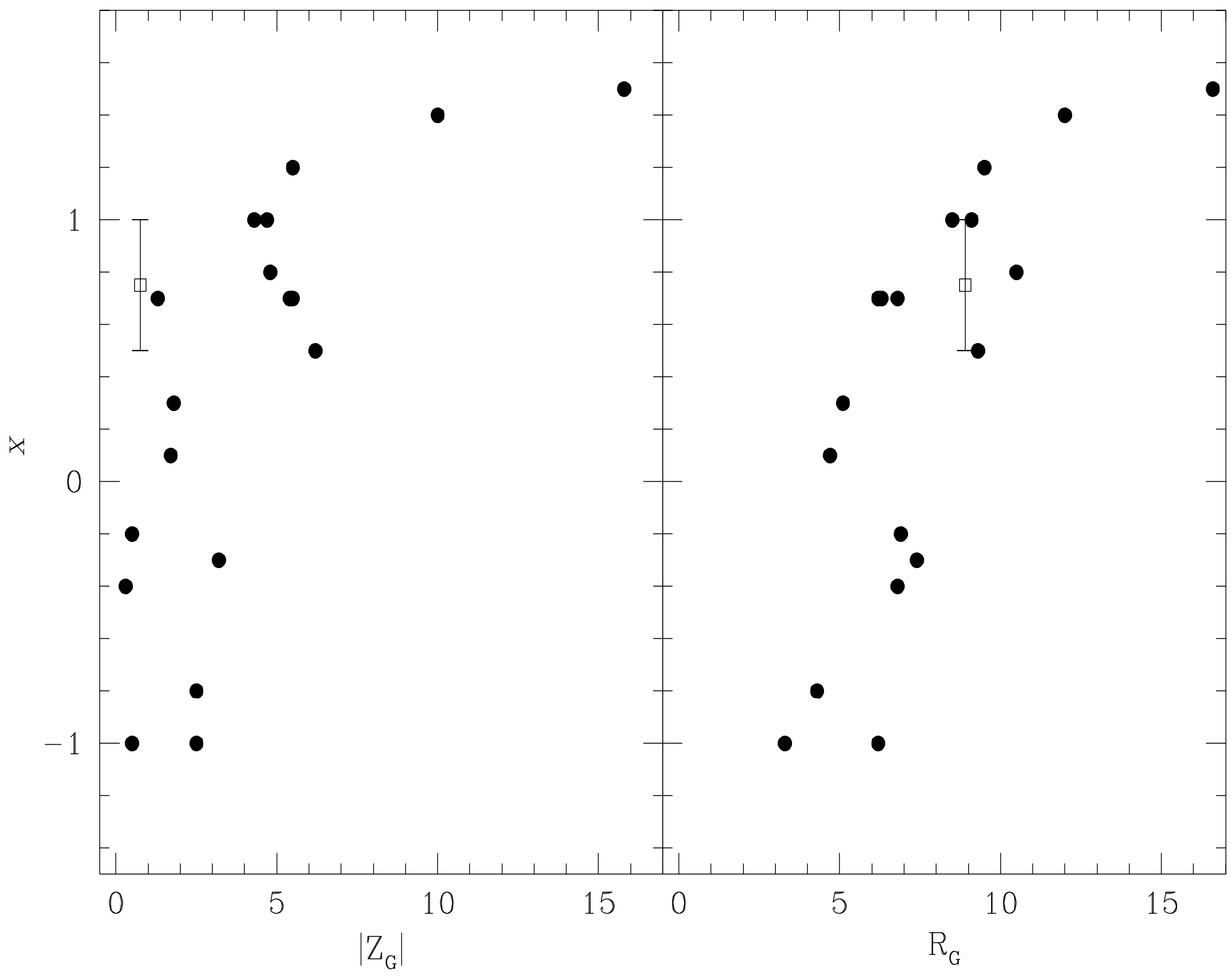

Figure 9

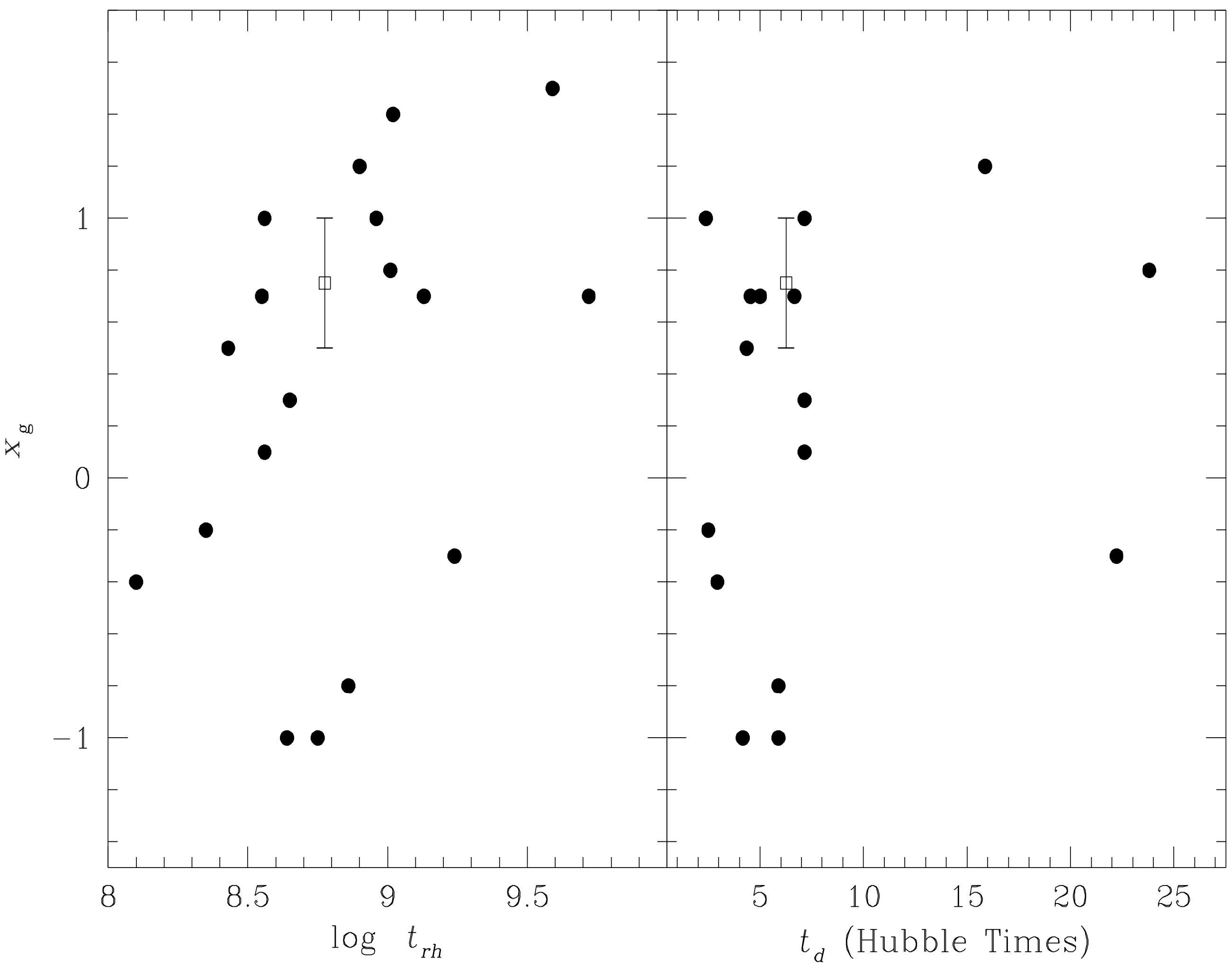

**Figure 10**